\documentclass{aa}  
\usepackage{threeparttable}
\usepackage{natbib}
\usepackage{xcolor}

\usepackage{soul}
\setstcolor{red}
\usepackage{float}
\usepackage{hyperref}
\usepackage{graphicx}
\usepackage{txfonts}
\usepackage{microtype}
\usepackage{placeins}

\begin{document}

   \title{A metal-poor ultra compact dwarf galaxy at a kiloparsec distance from the low-mass elliptical galaxy FCC\,47\thanks{Based on observation collected at the ESO Paranal La Silla Observatory,
Chile, Prog. ID 60.A-9192, PI Fahrion}}

\titlerunning{A metal-poor UCD at kiloparsec distance from FCC\,47}

   \author{Katja Fahrion
          \inst{1}
          \and
          Iskren Georgiev\inst{2}
          \and
          Michael Hilker\inst{1}
          \and
          Mariya Lyubenova\inst{1}
          \and
          Glenn van de Ven\inst{3}
          \and
          Mayte Alfaro-Cuello\inst{2}
          \and
           Enrico M. Corsini\inst{4}\fnmsep\inst{5}
           \and
           Marc Sarzi\inst{6}\fnmsep\inst{7}
           \and
          Richard M. McDermid\inst{8}
          \and
          Tim de Zeeuw\inst{9}\fnmsep\inst{10}}

   \institute{European Southern Observatory, Karl Schwarzschild Stra\ss{}e 2, 85748 Garching bei M\"unchen, Germany\\
   				\email{kfahrion@eso.org}
         \and
             Max-Planck-Institut f\"ur Astronomie, K\"onigstuhl 17, 69117 Heidelberg, Germany
             \and
             Department of Astrophysics, University of Vienna, T\"urkenschanzstrasse 17, 1180 Wien, Austria
             \and
             Dipartimento di Fisica e Astronomia ‘G. Galilei’, Università di Padova, vicolo dell’Osservatorio 3, I-35122 Padova, Italy
             \and
             INAF–Osservatorio Astronomico di Padova, vicolo dell’Osservatorio 5, 35122 Padova, Italy
             \and
             Armagh Observatory and Planetarium, College Hill, Armagh, BT61 9DG, UK
             \and
             Centre for Astrophysics Research, University of Hertfordshire, College Lane, Hatfield AL10 9AB, UK
             \and
             Department of Physics and Astronomy, Macquarie University, North Ryde, NSW 2109, Australia
             \and
             Sterrewacht Leiden, Leiden University, Postbus 9513, 2300 RA Leiden, The Netherlands
             \and
             Max-Planck-Institut f\"ur extraterrestrische Physik, Gie\ss{}enbachstraße 1, 85748 Garching bei M\"unchen, Germany
             }

   \date{}

% 5 {} token are mandatory
 
  \abstract
   {Photometric surveys of galaxy clusters have revealed a large number of ultra compact dwarfs (UCDs) around predominantly massive elliptical galaxies. Their origin is still debated as some UCDs are considered to be the remnant nuclei of stripped dwarf galaxies while others seem to mark the high-mass end of the star cluster population.}
   {We aim to characterise the properties of a UCD found at very close projected distance ($r_{\rm proj}$ = 1.1\,kpc) from the centre of the low-mass ($M \sim 10^{10}M_\odot$) early-type galaxy FCC\,47. This is a serendipitous discovery from MUSE adaptive optics science verification data. We explore the potential origin of this UCD as either a massive cluster or the remnant nucleus of a dissolved galaxy.}
   {We use archival \textit{Hubble Space Telescope} data to study the photometric and structural properties of FCC\,47-UCD1. In the MUSE data, the UCD is unresolved, but we use its spectrum to determine the radial velocity and metallicity.}
   {FCC\,47-UCD1's surface brightness is best described by a single King profile with low concentration $C = R_\text{t}/R_\text{c} \sim 10$ and large effective radius ($r_{\rm eff} = 24$\,pc). Its integrated magnitude and a blue colour ($M_g = {-10.55}$\,mag, $(g - z)$ = 1.46 mag) combined with with a metallicity of \mbox{[M/H] = $-$1.12 $\pm$ 0.10 dex} and an age $>$ 8 Gyr obtained from the full fitting of the MUSE spectrum suggests a stellar population mass of $M_\ast = 4.87 \times 10^6 M_\sun$. The low S/N of the MUSE spectrum prevents detailed stellar population analysis. Due to the limited spectral resolution of MUSE, we can only give an upper limit on the velocity dispersion ($\sigma < 17$ \,$\text{km s}^{-1}$), and consequently on its dynamical mass  ($M_\text{dyn} < 1.3 \times 10^7 M_\sun$).}
  {The origin of the UCD cannot be constrained with certainty. The low metallicity, old age and magnitude are consistent with a star cluster origin, whereas the extended size and high mass are consistent with an origin as the stripped nucleus of a dwarf galaxy with a initial stellar mass of a few $10^8 M_\sun$.}

   \maketitle

\section{Introduction}
Ultra compact dwarf galaxies (UCDs) were discovered about two decades ago \citep{Minniti1998, Hilker1999b, Drinkwater2000} in studies of galaxy clusters. With masses of \mbox{$10^6 < M < 10^8\,M_\sun$} \citep{Mieske2013} and radii between 10 and 100 pc \citep{Drinkwater2003, Hasegan2005, Mieske2006}, UCDs are among the densest stellar systems in the Universe. Compared to other dense systems, they occupy the region between globular clusters (GCs) and compact ellipticals (cEs) on the mass-size plane (e.g. \citealt{Misgeld2011, Brodie2011, Norris2014, Janz2016}). The intermediate nature of these objects reflects in the proposed formation scenarios for UCDs: i) UCDs constitute the high-mass end of classical GCs (e.g. \citealt{Mieske2002, Mieske2004, Kissler-Patig2006}), ii) are formed from the merging of star clusters in interacting galaxies \citep{Fellhauer2002, Maraston2004, Fellhauer2005} or iii) they are the remant nuclei of tidally stripped dwarf or low-mass galaxies \citep{Phillipps2001, Bekki2003, Drinkwater2003, Pfeffer2013, Strader2013}. While many UCDs occupy a similar parameter space as nuclear star clusters (NSCs) (e.g. \citealt{Walcher2005, Norris2015}), there is evidence that the general UCD population contains both genuine star clusters and remnant nuclei \citep{Brodie2011, DaRocha2011, Janz2016}. Very recently, \cite{Du2018} suggested the formation of compact ellipticals (cEs) through ram-pressure stripping induced star formation of a stripped, gas-rich dwarf galaxy that orbits in the hot corona of a massive galaxy. Although cEs are generally more massive and larger than UCDs, a similar formation scenario could apply to the most massive, metal-rich UCDs.

In the stripped nucleus scenario, tidal tails and extended envelopes around UCDs are expected. These are challenging to observe due to their short lifetimes; however, such features have been detected around some UCDs in the Fornax cluster \citep{Voggel2016, Wittmann2016}. In addition, the massive ($M = 4 \times 10^8 M_\sun$) second nucleus of the merger galaxy NGC\,7727, that probably resembles a UCD in formation, seems to be embedded in a tidal stream \citep{Schweizer2018}. If UCDs are remnant nuclei of stripped galaxies, one expects to find a massive black hole (BH) in the UCD's centre, similar to those found in some NSCs (e.g. \citealt{Seth2008, GrahamSpitler2009, Neumayer2012}). The presence of the BH causes a rise in the central velocity dispersion of the UCD, observable with high angular resolution integral field unit (IFU) instruments supported by adaptive optics (AO). Recent studies of massive UCDs in the Virgo and Fornax cluster \citep{Seth2014, Ahn2017, Ahn2018, Afanasiev2018} have indeed detected signatures of central super massive BHs (SMBHs) that constitute up to 15 \% of the UCD's mass. 

The presence of a central SMBH has also shed light on a interesting observational result from a decade ago. Based on measurements of structural parameters from photometry in combination to integrated velocity dispersion measurements, it was shown that many UCDs exhibit an elevated dynamical mass-to-light ratio ($M/L_\mathrm{dyn}$) when compared to stellar population estimates (e.g. \citealt{Hasegan2005, Mieske2008, Taylor2010}). Originally, this trend prompted suggestions that there is a variation of the initial mass function (IMF) in UCDs. Both, top-heavy \citep{Murray2009, Dabringhausen2009} and bottom-heavy IMFs \citep{MieskeKroupa2008} are discussed. 
While a SMBH presents a different explanation for the elevated $M/L_\mathrm{dyn}$ for high mass UCDs \citep{Mieske2013}, a search for SMBHs in two lower-mass UCDs ($< 10^7 M_\sun$) around Centaurus\,A yielded a non-detection \citep{Voggel2018}. Because of detection thresholds, the non detection of a central BH; however, does not disqualify a UCD from being a former nucleus of an accreted low-mass dwarf galaxy. 

Nonetheless, this non-detection might imply that the population of high-mass UCDs (above 10$^7 M_\sun$) are dominated by tidally stripped nuclei while many lower-mass UCDs are high-mass star clusters. This is also supported by semi analytical analyses of $\Lambda$CDM simulations \citep{Pfeffer2014, Pfeffer2016} that predicted the number and mass distribution of stripped nuclei in a cluster environment similar to the Virgo and Fornax clusters. 
As of now, the sample of identified NSC-type UCDs is biased to high stellar masses, where SMBHs can be found with high angular resolution IFU observations or the brightness of the object allows a detailed study of the star formation history (SFH) \citep{Norris2015}. These high-mass UCDs typically were once the nuclei of stripped massive galaxies with initial masses $> 10^{9} M_\sun$ \citep{NorrisKannappan2011, Seth2014, Ahn2018,  Afanasiev2018}. Nonetheless, the remnant nuclei of stripped  dwarf galaxies ($M_\ast \sim 10^8 M_\sun$) should be found among the low-mass UCDs ({$< 10^7 M_\sun$}), as e.g., $N$-body simulations show \citep{Pfeffer2013}; however, this sample is contaminated by high-mass GCs.

Observationally, it is difficult to differentiate what fraction of UCDs at each mass bin are stripped nuclei or are formed as genuine (or merged) star clusters at these low masses. Generally, NSCs of dwarf galaxies have larger sizes and are more massive and brighter than typical GCs \citep{Misgeld2011, Sandoval2015}. SMBHs in UCDs would give direct evidence but are very difficult to measure in distant low-mass systems \citep{Voggel2018}. In addition, the nuclei of dwarf galaxies are metal-poorer than their high-mass equivalents as dictated by the mass-metallicity relation \citep{Spengler2017} and thus have metallicities and colours comparable to the blue population of a GC system \citep{NorrisKannappan2011, Sandoval2015, Janz2016}. An extended or chemically complex SFH could give evidence for a NSC-origin (e.g. $\omega$Cen, \citealt{HilkerRichtler2000}); but studying SFHs of unresolved extragalactic UCDs requires deep spectroscopic observations \citep{NorrisKannappan2011} that are challenging for distant low mass systems.

In the Milky Way (MW), stripped nuclei of dwarf galaxies can be identified more easily. $\omega$Cen (NGC\,5139), the most massive GC in the MW \citep{Harris1996} with a mass of $\sim 3 \times 10^6 M_\sun$ \citep{Baumgardt2017, BaumgardtHilker2018}, is often considered to be the remnant nucleus of an accreted dwarf galaxies due to the presence of multiple stellar populations (e.g. \citealt{King2012}) and its retrograde orbit \citep{Majewski2000}. The second most massive star cluster M54 (NGC 6715, $M \sim 1.4 \times 10^6 M_\sun$, \citealt{BaumgardtHilker2018}) is classified as a GC of the MW \citep{Harris1996}; however, it is located in the centre of the Sagittarius dwarf galaxy (e.g. \citealt{Ibata1997, Bellazzini2008}) and thus is considered to be the NSC in the process of being stripped by the MW. Both $\omega$Cen and M54 are metal-poor ([Fe/H] = -1.62 dex and -1.58 dex, respectively) and have absolute magnitudes of $M_V \sim -10.37$ mag and $-10.20$ mag, respectively \citep{Harris1996}.

In addition to observational challenges to identify low-mass UCDs among the GC population of a galaxy, it is unclear if the formation pathway of UCDs depends on the galaxy cluster environment they reside in. 
Typically, UCDs are found at rather large projected distances ($>$ 10 kpc) from a nearby giant elliptical galaxy located near the centre of a galaxy cluster or group. To name one, the first discovered UCD in \cite{Hilker1999b}; named UCD3 in \citealt{Drinkwater2000}, has a projected distance of 11 kpc to the closest galaxy NGC 1404, a giant elliptical galaxy in the centre of the Fornax cluster. However, its radial velocity suggests that UCD3 is instead bound to NGC 1399, located 50 kpc in projected distance from UCD3. It is not clear whether these large separations are connected to the formation of UCDs or whether they are a selection effect as it is difficult to perform a systematic search for UCDs in the high surface brightness regions near the centre of major galaxies. 

\begin{figure}
\centering
\includegraphics[width=0.49\textwidth]{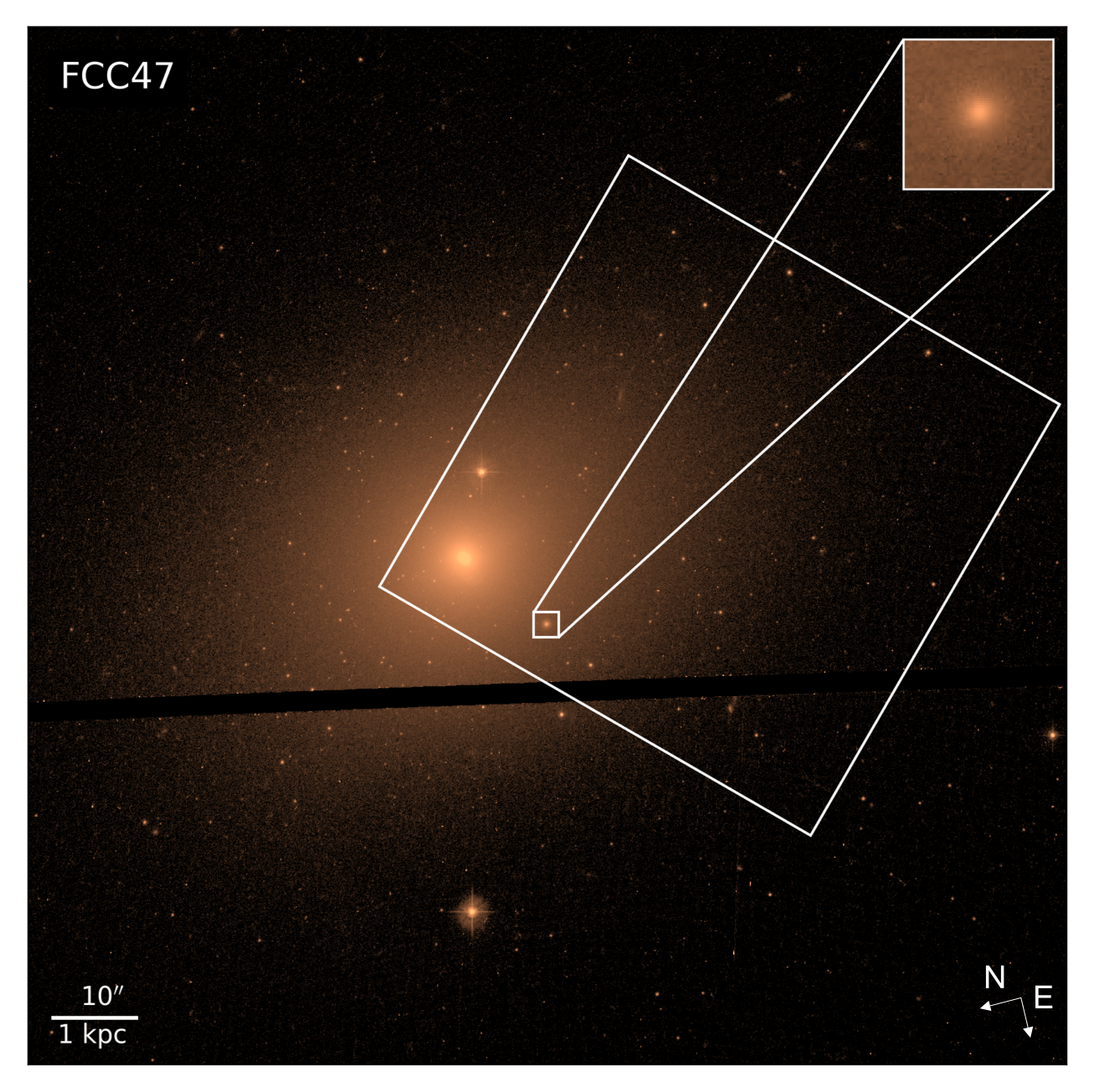}
\caption{FCC\,47 HST/ACS F475W image of the central $\sim$ 11 $\times$ 11 kpc. The UCD is shown in the zoomed box. The white rectangle shows the pointing of the MUSE field-of-view in wide field mode corresponding to 1\arcmin $\times$ 1\arcmin.} 
\label{fig:Image}
\end{figure}

In this paper we report the serendipitous discovery of a UCD in close projected distance (r$_\text{proj}$ = 1.1 kpc) to the low-mass early-type galaxy FCC\,47 (NGC\,1336) in Multi Unit Spectroscopic Explorer (MUSE) adaptive optics (AO) science verification (SV) data. For the photometric analysis, we use archival \textit{Hubble Space Telescope} (HST) data taken with the Advanced Camera for Surveys (ACS). A zoomed HST/ACS image in the F475W filter of FCC\,47 is shown in Figure \ref{fig:Image}. The white contours show the placing of the MUSE pointing. It is oriented such that the NSC of FCC\,47 - the main target of the original study - is in the corner. With a projected distance of 780 kpc to NGC 1399, FCC\,47 itself lies in the outskirts of the Fornax cluster, but was covered by photometric surveys of the cluster such as the HST ACS Fornax Cluster survey (ACSFCS, \citealt{Jordan2007}) and the Fornax Deep Survey (FDS, \citealt{Iodice2016}). However, the UCD in FCC\,47 (hereafter FCC\,47-UCD1) has not been discovered in these surveys, probably because their method of selecting classical GCs missed this UCD due to its large size and extended nature. Table \ref{tab:pointing} lists some basic information about FCC\,47. We explore the various observational parameters of FCC\,47-UCD1 to constrain the possible origin of the UCD as either a massive GC or the remnant NSC of a stripped dwarf galaxy.

This paper is organized as follows: Section \ref{sect:data} presents the data. Section \ref{sect:methods} contains our structural photometric and spectroscopic analysis. The results are shown in Section \ref{sect:results} and are discussed in Section \ref{sect:discussion}. In Section \ref{sect:conclusion}, we present a summary and the conclusions in Section \ref{sect:conclusion}.

%--------------------------------------------------------- HST F475W with MUSE contours 
   
\begin{table}
    \centering
    \caption{Basic information about FCC\,47.}
    \begin{threeparttable}
    \begin{tabular}{ l  c  c }\hline \hline
    Property & Value & Reference \\ \hline
  RA (J2000)		& 03:26:32.19 & \\
  DEC  (J2000)	&  -35:42:48.80 & \\
 d [Mpc] & 18.3 $\pm$ 0.6 & \cite{Blakeslee2009} \\
 B$_T$ [mag] & 13.34 & \cite{Glass2011} \\
 $r_\text{eff}$ [arcsec] & 30.0 & \cite{Ferguson1989} \\
 $M_\ast$ [$M_\sun$] & $\sim 2 \times 10^{10}$ & Fahrion et al. (in prep)\tnote{a} \\ \hline
\end{tabular}
\begin{tablenotes}
\item[a] Based on orbit-based dynamical model. \cite{Saulder2016} report a stellar mass of 6.4 $\times 10^9 M_\sun$.
\end{tablenotes}
    \label{tab:pointing}

    \end{threeparttable}
\end{table}
%-----------------------------------------------------------------------------%-----------------------------------------------------------------------------
%-----------------------------------------------------------------------------%-----------------------------------------------------------------------------
%%% Data: HST + MUSE data
%% MUSE data: upcoming paper?

\section{Data}
\label{sect:data}
\subsection{MUSE SV AO data}
\label{subsect:MUSE_data}
We discovered FCC\,47-UCD1 in AO-supported IFU data of FCC\,47 acquired with the MUSE instrument on the Very Large Telescope. In the wide field mode (WFM, \citealt{Bacon2010}) MUSE provides $\sim$ 90\,000 spectra in an almost square 1\arcmin$\times$1\arcmin\, field-of-view (FOV) with a spatial sampling of 0.2\arcsec/spaxel. MUSE operates in the optical wavelength regime with a mean spectral resolution of R $\sim$ 3000 (d$V \sim 100$ km s$^{-1}$). Our data were acquired during the SV phase (programme 60.A-9192, P.I. Fahrion) following the commissioning of the Ground Atmospheric Layer Adaptive Corrector for Spectroscopic Imaging (GALACSI) AO system for the WFM. The data were acquired on the 17th of September 2017. The observing program was designed to study the NSC of FCC\,47 in combination with FCC\,47's rich GC system \citep{Jordan2015}, which will be presented in a forthcoming paper (Fahrion et. al. in prep). The simultaneous observation of the UCD in the MUSE FOV was purely coincidental. 

The MUSE data of FCC\,47 consist of ten exposures of 360 seconds each. Dedicated sky exposures of three minutes were acquired in between the on-source observations. 
The data were taken in the nominal mode with a wavelength coverage ranging from 4750 to 9300 \AA. Due to the sodium lasers of the AO facility, the wavelength range between 5800 and 6000 \AA\, is filtered out. The observing conditions for FCC\,47 were less than ideal with an atmospheric seeing of around 1.6\arcsec. However, the GALACSI system was capable to reduce this to a final point spread function (PSF) with a full width at half maximum (FWHM) of $\sim$ 0.7\arcsec. 
Unfortunately, because there is only one bright foreground star in the FOV and it is located in the high surface brightness region near the centre of the galaxy, the data are not ideal to study the effects of the AO system on the MUSE PSF. We measured the PSF FWHM on the star with \textsc{iraf} and by two-dimensional (2D) modelling with \textsc{imfit} \citep{Erwin2015}. Both, a Gaussian model and a Moffat profile fit the PSF equally well, so for the purpose of this work we assume a Gaussian PSF with a FWHM of $\sim$ 0.7\arcsec. At this image quality, the UCD is not spatially resolved in the MUSE data, but is visible as a bright point source in the high surface brightness region of the galaxy.

The FCC\,47 MUSE data were reduced using the MUSE data reduction pipeline version 2.2 \citep{Weilbacher2014, Weilbacher2016} incorporated in an ESO \textsc{Reflex} workflow \citep{Freudling2013}. The reduction follows the steps described in \cite{F3D_Survey}. It includes bias correction, flat-fielding and a wavelength calibration. The sky subtraction was done by subtracting a model sky spectrum acquired from fitting the dedicated sky exposures. 
To reduce the sky residuals further, we applied the the ZAP algorithm version 2.0 \citep{Soto2016}. This procedure uses principle component analysis to statistically quantify and then remove the sky line residuals.

\subsection{HST/ACS data}
\label{subsect:HST_data}
We complement the MUSE data with archival HST/ACS data in the F475W and F850LP filters (roughly Sloan's $g$ and $z$-bands) from the ACSFCS (PI: A. Jord\'an). The data are of high quality with 760 (F475W) and 1220 (F850LP) seconds of exposure time. There is also 1996-05-10 WFPC2 data in the F606W filter (HST proposal 5446 PI: G. Illingworth) that covers FCC\,47 and the UCD, but it has only 80 s of exposure time. Due to the low image quality, we do not use it for the analysis of the structural properties of the UCD in this filter, but list the integrated magnitudes in Table \ref{tab:UCD_params}.

We processed with \textsc{Astrodrizzle} the ACS .flc files (corrected already by the HST reduction pipeline for dark, bias, flat field and charge transfer efficiency), and the corresponding raw files for WFPC2. Because there were only two exposures per filter, we chose to drizzle the images to their native pixel scale of the detector of 0.05\arcsec pix$^{-1}$ for ACS and the WFPC2 PC chip, where the UCD is located. Because FCC\,47 is quite extended, we disabled \textsc{astrodrizzle} from automatically obtaining sky levels. Instead we derived our own sky values in a region without galaxy contribution and supplied them as input. The spatially variable PSF model for each ACS filter was built using around 10 foreground stars. Due to the small FOV of the WFPC2 PC-chip, there were no suitable stars from which to build a PSF. Therefore, we generated the PSF with \textsc{tinytim} for the location of the UCD on the WFPC2/PC chip and drizzled the PSF images with the same \textsc{astrodrizzle} parameters as for the science data. For the computation of magnitudes, we use photometric zeropoints in the Vega system of 26.172, 22.850 and 24.352 mag for the F475W, F606W and F850LP filters, respectively.

%--------------------------------------------------------------------
%--------------------------------------------------------------------%--------------------------------------------------------------------
\section{Analysis}
\label{sect:methods}
\subsection{Structure and photometry}
\label{subsect:photometry}
We study the photometry of FCC\,47-UCD1 in the HST filters F475W ($g$) and F850LP ($z$). As the UCD is close in projection (13\arcsec = 1.1\,kpc) to the centre of FCC\,47, the galaxy background has to be removed. This is achieved by creating an \textsc{iraf ellipse} model of the host galaxy in each filter and subtracting it from the data. We find that the residual is flat at the UCD's position. The residual varies with a standard deviation of 6.4 counts s$^{-1}$ around a mean flux level of 3.7 counts s$^{-1}$ in the F475W filter, whereas the UCD reaches a peak of 1245 counts s$^{-1}$, corresponding to a magnitude fluctuation $\sim 0.005$ mag. The structural parameters of the UCD are then determined using \textsc{imfit} \citep{Erwin2015}, a modular procedure to fit 2D surface brightness distributions. 

\textsc{imfit} offers a variety of 2D surface brightness models that can be added freely to create multi-component or off-center component models. In the past, UCD surface brightness profiles have been successfully fitted by either King or S\'{e}rsic profiles (e.g. \citealt{Hasegan2005, Evstigneeva2007, Mieske2008}), often also using two component models representing a dense core and a more diffuse stellar envelope. We use \textsc{imfit} to fit FCC\,47-UCD1 with two single component models (King and S\'ersic, respectively) and two two-component models (King + Exponential and King + S\'ersic). We use generalized King models\footnote{The generalized King profile is given by \\ $I(R) = I_0 \left[\frac{1}{(1 + (R/R_\text{c})^2)^{\frac{1}{\alpha}}} - \frac{1}{(1 + (R_\text{t}/R_\text{c})^2)^{\frac{1}{\alpha}}}\right]^\alpha$} that are parametrised by a core radius $R_\text{c}$, a concentration $C = \text{R}_t/\text{R}_c$ as ratio between tidal and core radius and a variable exponent $\alpha$. We use 10$\times$ oversampled PSFs in the two filters to ensure accurate representation of the smallest scales.
\textsc{imfit} allows to determine the best-fitting parameters with a Markov-Chain-Monte-Carlo (MCMC) analysis. The MCMC analysis of our fits shows well behaved distributions around the best-fit parameters for the single component fits (see Figure \ref{fig:corner}). For the two component fits we see degeneracies between many parameters indicating that these models over-fit the data. 

\begin{figure*}[h!]
\centering
\includegraphics[width=0.98\textwidth]{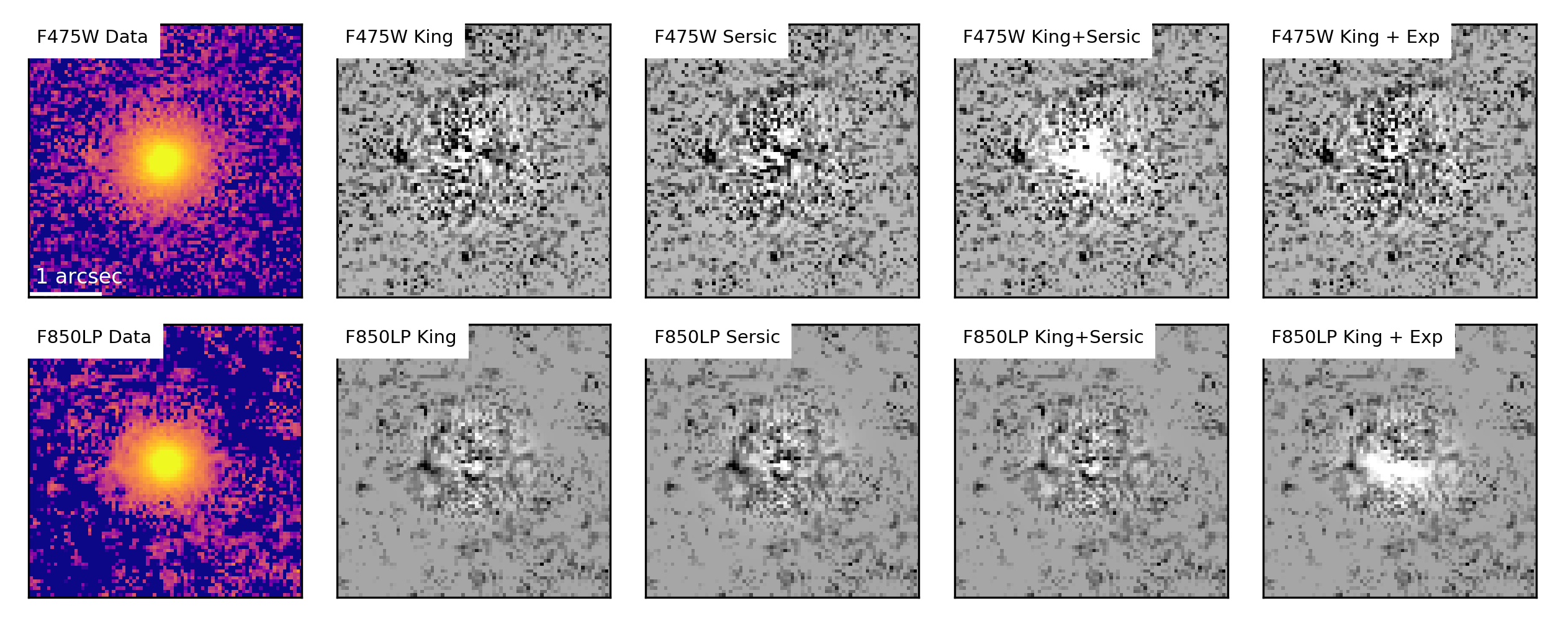}
\caption{2D structural analysis. \textit{First column}: HST UCD images in the F475W filter ($g$-band, \textit{first row}) and F850LP ($z$-band, \textit{second row}) after an \textsc{ellipse} model of the galaxy background is subtracted. The second and third columns show residual images after subtracting a single generalized King model or a S\'ersic model, respectively. The third and forth columns show residuals after subtracting two component models with a King and S\'ersic and a King and exponential model, respectively. Each panel shows a $3\arcsec\times3\arcsec$ (266 $\times\,266$ pc) cut out. Black and white colours show positive and negative residuals. See Figure \ref{fig:profiles} for radial relative residuals from the various models.}
\label{fig:2D_plot}
\end{figure*}

\begin{figure}
\centering
\includegraphics[width=0.48\textwidth]{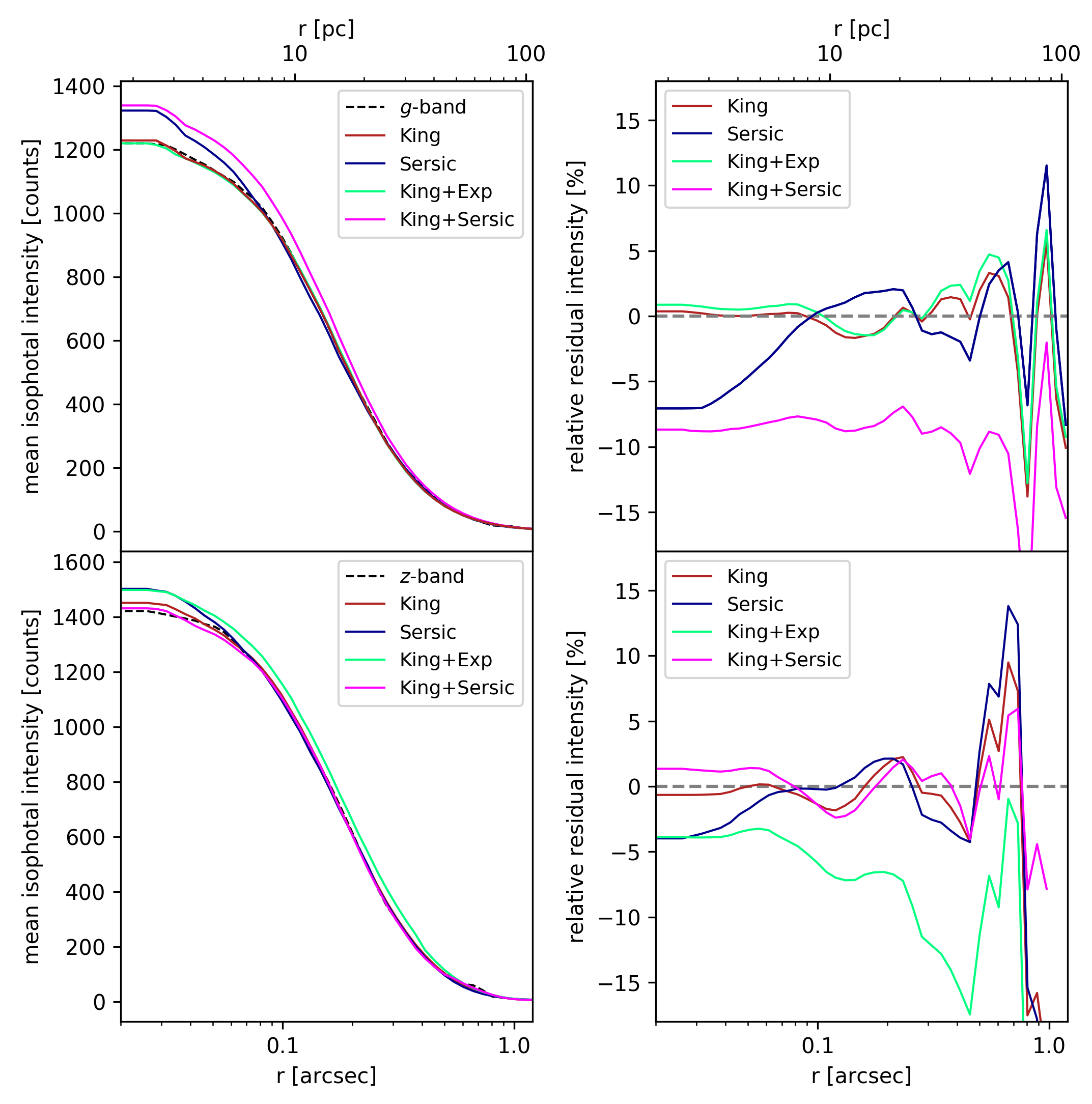}
\caption{\textit{Left column}: UCD 1D profiles of the mean isophotal intensity versus radius in the $g$-band (\textit{first row}) and $z$-band (\textit{second row}) filter. The data is shown with the black dotted line, the King model fit in red, the S\'ersic model fit in blue. The two component fits are shown in cyan (King + Exponential) and magenta (King + S\'ersic). \textit{Right column}: relative residual profiles for the four different models. Above $\gtrsim 5$\arcsec the flux drops to the background level, creating noisy radial residuals.} 
\label{fig:profiles}
\end{figure}

The 2D surface brightness images of the UCD and the residuals after subtracting the different \textsc{imfit} models are shown in Figure \ref{fig:2D_plot}. 
The single component fits perform well with no obvious difference between King and S\'{e}rsic models. In contrast, the two component models tend to overestimate the UCD flux at the centre. To further investigate the fit, we show the radial intensity profiles and the relative residuals extracted with the \textsc{iraf ellipse} task in Figure \ref{fig:profiles}. As seen in the 2D plot, the two component models overestimate the flux in the centre. In the $g$-band, the difference between King and S\'{e}rsic profile is small, but the single King profile performs slightly better, as quantified by the residuals (right panel). Similarly, the best fit to the $z$-band data is achieved with the single King profile. In both filters, the relative residuals of the single King fits are below 3\% (corresponding to a magnitude error of $\sim$ 0.03 mag) out to three core radii. Above $\gtrsim 0.5$\arcsec\, the flux drops to the background level of $\sim$25 mag arcsec$^{-2}$ creating residuals with higher relative amplitudes.

We conclude that FCC\,47-UCD1 is best described by a single generalized King profile with the parameters listed in Table \ref{tab:UCD_params}. The formal errors of the MCMC analysis are $< 1\%$. To get realistic errors, we use \textsc{imfit's} bootstrapping analysis with 1000 iterations on the MCMC parameters. The resulting histograms can be found in Figure \ref{fig:bootstrap}. The bootstrapping gives no direct uncertainty for the integrated magnitude, only for the peak intensity as a parameter of the model of $I_0 = 1778.9 \pm 19.1$ counts s$^{-1}$ in the F475W filter. This corresponds to a magnitude uncertainty of 0.01 mag. For comparison, we use \textsc{iraf's phot} task to extract magnitudes directly from the background subtracted data. Here, we obtain uncertainties of 0.007 mag for both filters, independent of the chosen aperture size. Using the original data, these uncertainties slightly increase to $\sim$ 0.009 mag because of the uneven galaxy background. Systematic effects, e.g, created by the galaxy background thus are very small and we find that using a median filter for the background subtraction yields the same results. We therefore adapt an magnitude uncertainty of 0.01 mag for the integrated magnitudes in the F475W and F850LP filters. For the WFCP2/PC F606W magnitudes, we obtain an uncertainty of 0.10 mag due to the lower image quality. Our \textsc{ellipse} residuals are flat on the extend of the UCD and should not introduce an magnitude error larger than $0.005$ mag, but we tested other methods for the background subtraction such as using a simple median filter and found no difference in the results.
For the comparative analysis we transformed the ACS/WFC F475W and the WFCP2/PC F606W magnitudes to Johnson $B$ and $V$ magnitudes using, respectively, the \cite{Sirianni2005} and \cite{Dolphin2009} transformations. We obtain $V$ = 20.37 $\pm 0.10$ mag and $B$ = 21.09 $\pm$ 0.10 mag, corresponding to $M_V = -11.15 \pm 0.10$ mag using a distance of 18.3 Mpc \citep{Blakeslee2009}. The uncertainties are dominated by the photometric uncertainty of the F606W magnitude.

%-------------------------------------- 2D Plot

\begin{table*}
\centering
\caption{Photometric  and structural parameters of FCC\,47-UCD1.}
\begin{threeparttable}
\begin{tabular}{c c c c c c c c c}
\hline \hline
Filter & $m_0$ & $M$ & $\epsilon$ & \multicolumn{2}{c}{$R_{\rm c}$} & $C$ & $\alpha$  \\
	& [mag] & [mag] &  & [arcsec] &  [pc] &  &  \\
(1)    & (2)     & (3)    & (4) &  \multicolumn{2}{c}{(5)}    & (6) & (7) \\ \hline   
F475W  ($g$) & 20.76 $\pm$ 0.01  &  $-$10.55 $\pm$ 0.01 & 0.08 $\pm$ 0.01 & 0.129  $\pm$ 0.002 & 11.4 $\pm$ 0.2 & 12.23 $\pm$ 0.47 & 1.69 $\pm$	0.07 \\
F606W & 20.14 $\pm$ 0.10 & -11.26 $\pm$ 0.10 & \\
F850LP ($z$) & 19.30  $\pm$ 0.01 & $-$12.01 $\pm$ 0.01 & 0.08 $\pm$ 0.01 &  0.124 $\pm$ 0.002 & 11.0 $\pm$ 0.2 & 8.96 $\pm$ 0.34  & 1.37 $\pm$ 0.10  \\ \hline
\end{tabular}
\begin{tablenotes}
\item (1) HST filter, (2), (3) extinction-corrected apparent and absolute King model magnitude in the Vega system. We use $A_{\text{F475W}} = 0.04$, $A_{\text{F606W}} = 0.03$ and $A_{\text{F475W}} = 0.01$. (4) ellipticity, (5) core radius of the King model, (6) concentration parameter as the ratio between tidal and core radius, (7) exponent $\alpha$ of the King model. The uncertainties were obtained from bootstrapping with \textsc{imfit}, see Figure \ref{fig:bootstrap}. The effective radius in 2D and 3D as obtained from our mass modelling method are listed in Table \ref{tab:UCD_mass_modelling}.
\end{tablenotes}
\end{threeparttable}
\label{tab:UCD_params}
\end{table*}

% reff = 0.547 c ** 0.486

%-------------------------------------- fitted profiles

%-------------------- MUSE SPECTRUM -------------------------------------------------------------------------
\subsection{MUSE spectrum}
\label{subsect:MUSE_spec}
In the MUSE data, the UCD is not resolved, but it is clearly visible as a bright point source in the FOV. While it was overlooked in the HST/ACS study of GC candidates \citep{Jordan2015} probably due to its extended size, we can use the MUSE spectrum to confirm its membership to the FCC\,47 system and can further analyse the integrated kinematics and stellar population properties.

We extract the MUSE spectrum of the UCD in a circular aperture with 10 pixel radius and apply PSF weighting with a FWHM of 0.7\arcsec and a Gaussian shape. The PSF weighting is used to maximise the contribution from the UCD to the spectrum while simultaneously minimizing the influence of the strong underlying galaxy background. Since this spectrum is still heavily contaminated by the light from FCC\,47's stellar body, we extract the spectrum of the local galaxy background in an annular aperture placed around the UCD with 5 pixel width and 8 pixel radius from the UCD. The background-subtracted UCD spectrum has an average signal-to-noise ratio (S/N) of $\sim20$ \AA$^{-1}$ measured in a continuum region around 6500 \AA. This S/N is sufficient to safely derive the radial velocity $v_\text{rad}$ and, in principle, sufficient to determine the velocity dispersion. However, for such a low-mass system as the UCD, the velocity dispersion is close to or below the MUSE instrumental resolution as we discuss below.
We also obtain the mean spectroscopic metallicity of the UCD and estimate the age. The star-formation history or alpha-enhancement are also not accessible at this low S/N. 

We use the Penalized Pixel-Fitting code \textsc{pPXF} \citep{Cappellari2004, Cappellari2017} to extract the integrated kinematics and stellar population properties such as metallicity and age from the UCD spectrum after subtracting the galaxy contribution. \textsc{pPXF} is a full-spectrum fitting method that uses a penalized maximum likelihood approach to fit a spectrum with a combination of template spectra. We use the single stellar population (SSP) template spectra from the extended Medium resolution INT Library of Empirical Spectra (E-MILES, \citealt{Vazdekis2012, Ricciardelli2012, Vazdekis2016}) because of its broad wavelength coverage. The template spectra cover a range from 1680 to 50000 \AA, with a spectral resolution of 2.5 \AA, the mean instrumental resolution of MUSE. 
We use the  base [$\alpha$/Fe] models with BaSTI isochrones \citep{Pietrinferni2004, Pietrinferni2006} and a double power law (bimodal) IMF with a high mass slope of 1.30. The library offers 636 spectra in total, with 12 different metallicities from [M/H]$= -$2.27 to 0.4 dex and 53 ages from 30 Myr to 14 Gyr. The library does not offer alpha-enhanced models at this wavelength range which might result in slight overestimation of the metallicity if the UCD is alpha-enhanced.

Performing an unregularized fit with the full grid of E-MILES SSP templates we find a mean age of 13 Gyr with no weights in models with ages below 11 Gyr and a mean metallicity of \mbox{$\sim$ $-$1.1\,dex}. While this age estimate is probably rather uncertain due to the low S/N, we can exclude a dominant contribution to the UCD's spectrum from young stellar populations ($<$ 8 Gyr). This already shows that FCC47-UCD1 is clearly dominated by old stellar populations; however, neither with the MUSE spectrum nor the information from the ($g$ - $z$) colour (see Section \ref{sect:metallicity}) we can constrain the age further.  In the following, we assume it to be $\sim$ 13 Gyr.

\begin{figure*}[h!]
\centering
\includegraphics[width=0.98\textwidth]{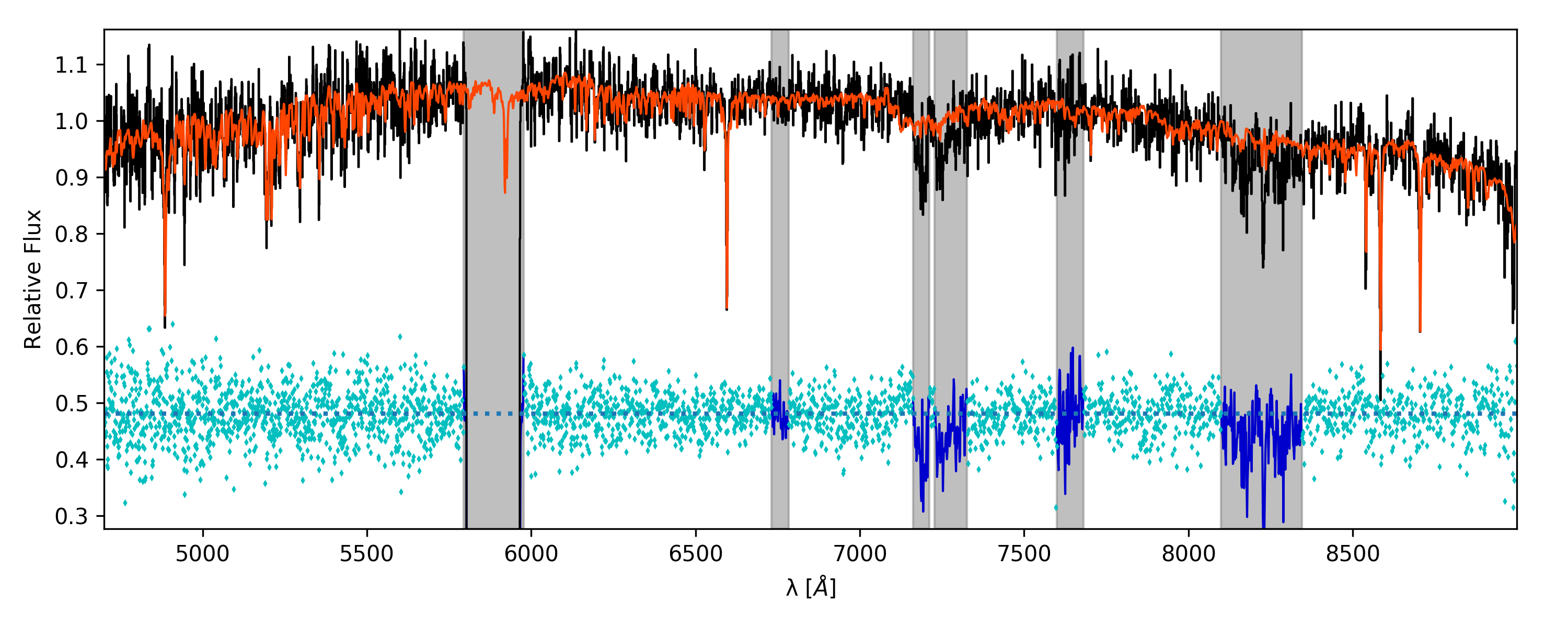}
\caption{\textsc{pPXF} fit to the UCD MUSE spectrum. The original spectrum is shown in black, the \textsc{pPXF} fit using the E-MILES templates \citep{Vazdekis2010} in red and the residual in blue. Masked regions with strong sky line residuals are shown in grey. The first masked region without spectrum is due to the AO laser filter.}
\label{fig:UCD_spec}
\end{figure*}

We determine the uncertainties on $v_\text{rad}$ and the mean metallicity [M/H] in a Monte Carlo (MC)-like approach: We perform a first fit and then create 600 realisations of the spectrum by adding the residual from the first fit in a random fashion to the best-fitting model. The perturbed spectrum is then fitted again. This way, we obtain a well-sampled distribution. In each fit, we first determine the kinematics with additive polynomials of degree 10 and then fit for the metallicity with multiplicative polynomials of degree 10 while leaving the kinematic parameters fixed. This way, we ensure that no additive polynomials are used in the stellar population fit that might change the line strengths. To avoid effects of of any age-metallicity degeneracy and to speed up the MC-runs, we restrict the template library to ages $>$ 8 Gyr.
Figure \ref{fig:UCD_spec} shows the original UCD spectrum with the best-fitting model overplotted in red. We show the distributions from the 600 MC runs in the Appendix (Figure \ref{fig:vs_metals_distributions}).

\begin{figure*}[h!]
\centering
\includegraphics[width=0.99\textwidth]{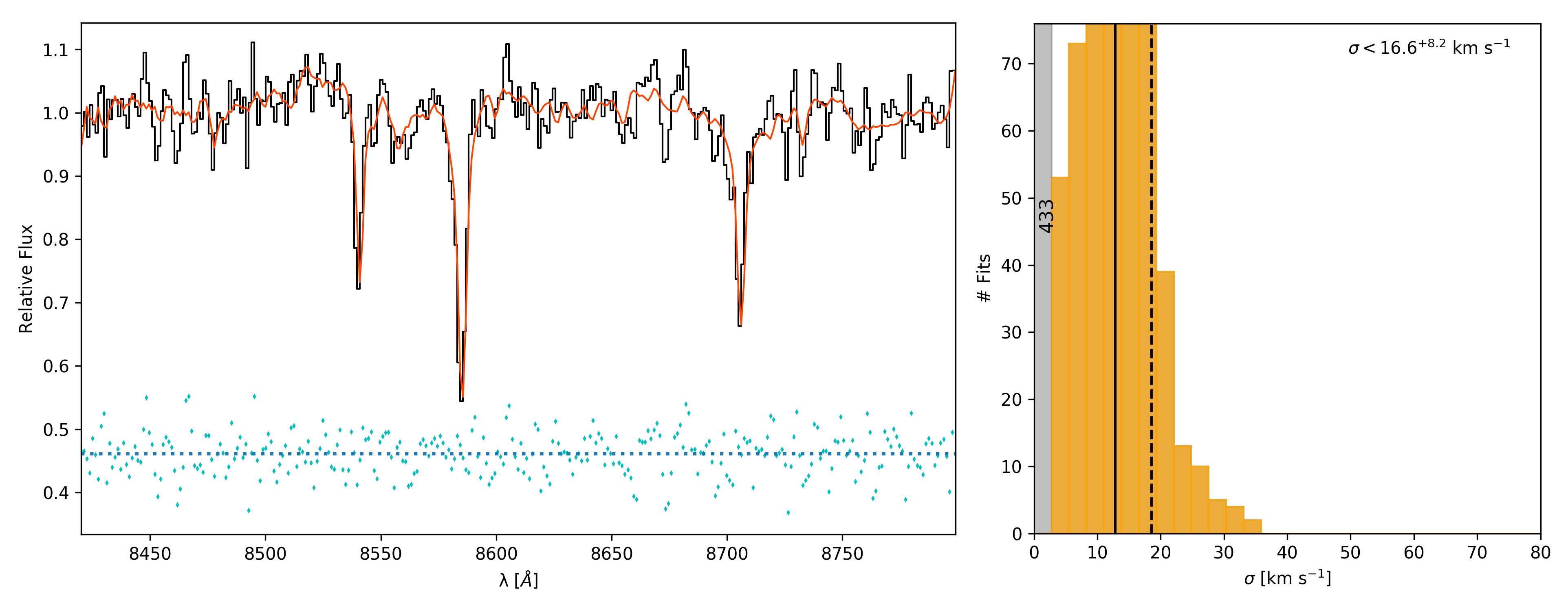}
\caption{\textit{Left}: Ca Triplet region of the MUSE spectrum. The \textsc{pPXF} fit using the Ca Triplet library shown in red. \textit{Right}: Distributions of velocity dispersions using a MC approach with 1000 realisations of the spectrum. 433 fits that returned zero are shown in the grey bar. The solid line gives the mean of the distribution after removing the runs that returned zero. The dotted line gives the upper 1$\sigma$ deviation.}
\label{fig:CaT}
\end{figure*}

Being a low-mass system, fitting the UCD's velocity dispersion from the background-subtracted spectrum is challenging due to the low S/N and the limited spectral resolution of the MUSE instrument. As the instrumental resolution is highest around the Calcium II triplet (CaT) at $\sim$8500 \AA\, \citep{Guerou2017}, we determine the velocity dispersion in a narrow wavelength range from 8420 to 8800 \AA. For this fit, we use the Calcium II Triplet library \citep{Cenarro2001}, consisting of $\sim$ 700 stars in a wavelength range between 8350 - 9020 \AA\ at 1.5 \AA\, spectral resolution. The higher resolution is relevant because the MUSE resolution as measured by \cite{Guerou2017} is below the resolution of the MILES templates of 2.5 \AA\, in the CaT region. This would require us to degrade the MUSE spectrum in this wavelength range and would further complicate the velocity dispersion measurement. We therefore chose the CaT library selecting F, G, K and M stars as representatives for old populations from the library and create 1000 realization as described above. This high number is needed because nearly half of all fits hit the lower limit given by \textsc{pPXF}. Removing these cases, we get the distribution shown in the right panel of Figure \ref{fig:CaT}. The left panel shows the MUSE spectrum around the CaT region and the \textsc{pPXF} fit.

%--------------------------------------------------------------------
%-------------------------------------------------------------------- RESULTs %--------------------------------------------------------------------
\section{Results}
\label{sect:results}
\subsection{Colour profile}

\begin{figure}
\centering
\includegraphics[width=0.48\textwidth]{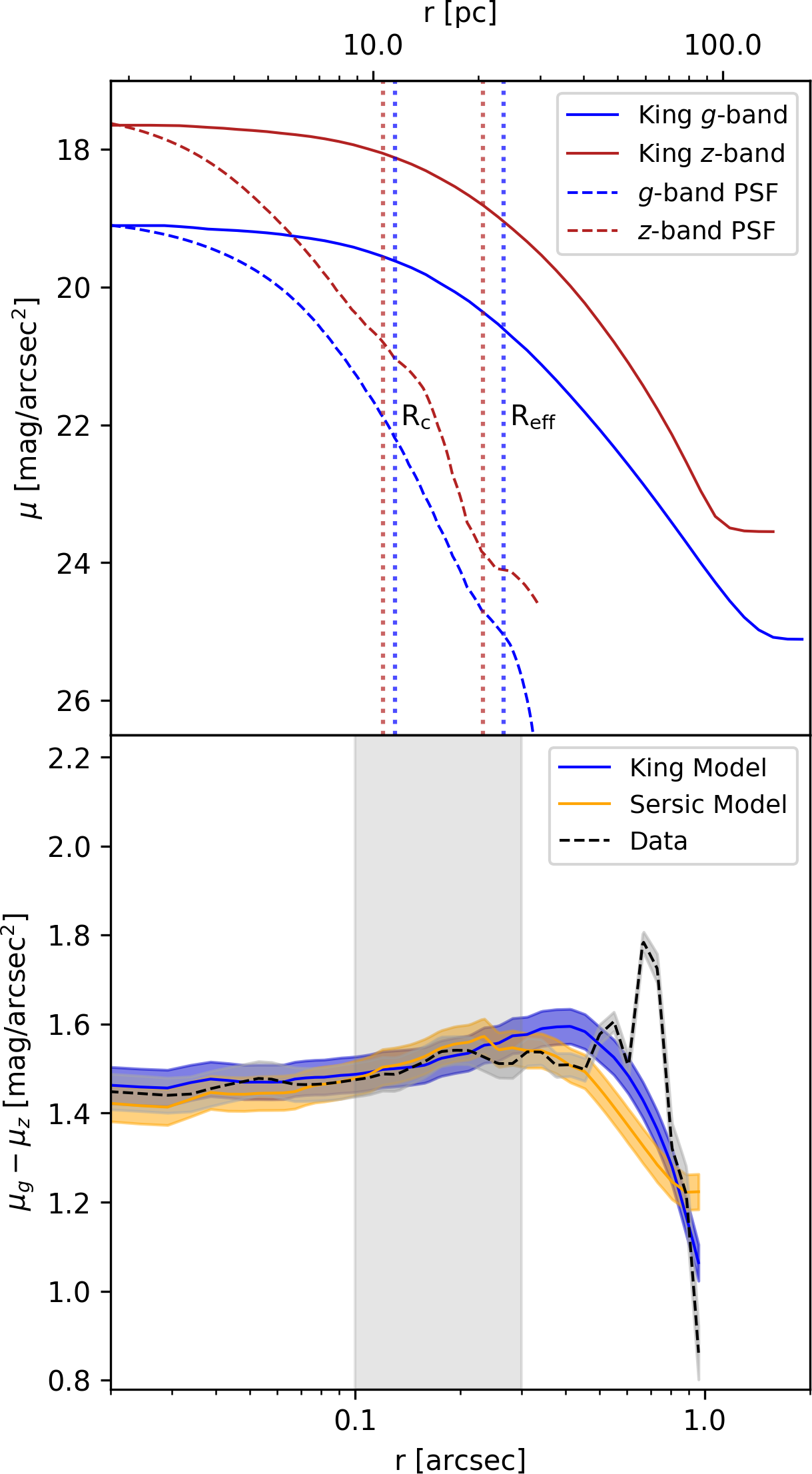}
\caption{Surface brightness and colour profiles. \textit{Top}: UCD model surface brightness profiles for the generalized King models. We also show the profiles from the 10$\times$ oversampled PSFs shifted to match the respective filter at the centre. The vertical lines give the positions of the core and effective radii in the two filters. \textit{Bottom}: $\mu_{g}-\mu_{z}$ colour profile. The blue line represents the King model colour profile, the black dotted line the data. For comparison, we also show the colour profile from the single component Sersic model. The shaded areas show the 1$\sigma$ uncertainties from the \textsc{ellipse} extraction and the grey-shaded area shows the range we use to fit the gradient.} 
\label{fig:color_profiles}
\end{figure}

The surface brightness profiles of the UCD are shown in Figure \ref{fig:color_profiles} together with its $(g - z)$ colour profile. The shaded areas refer to the errors from \textsc{ellipse}. The total colour in Vega magnitudes is $(g - z) = $ 1.46 mag. 
The PSF surface brightness profiles shown as dotted lines fall faster than the data and should not affect the colour profile above the core radius ($r\gtrsim0.1\arcsec\simeq2\times{\rm FWHM_{PSF}}$), where the PSFs are already 2.5 mag/${\rm arcsec}^2$ fainter than the UCD. The PSF profiles shown in this figure are taken from the 10$\times$ oversampled PSF models.
In the region, where we can exclude PSF effects and noise due to the low surface brightness in the outer parts (0.1 $< r < 0.3$\arcsec), the colour profile as extracted directly from the data of the UCD shows a mild color gradient of 0.17 mag. Using the generalized King model colour profile or as a comparison the single component Sersic model profile, the gradient is more prominent (0.4 mag) in this region. With our combined photometric uncertainties of 0.014 mag, this gradient is significant.

We also consider the influence of the galaxy at this position.
FCC\,47 is much redder than the UCD and has a colour of ($g - z$) $\approx$ 2 mag at the UCD's position. 
This could theoretically bias the UCD colour profile; however, the galaxy surface brightness at that location is two magnitudes fainter than that of the UCD. It should therefore only account for $\sim$ 16\% of the flux, if we did not subtract the \textsc{ellipse} model. As our \textsc{ellipse} residual introduces an error below 0.01 mag, we are confident that the UCD colour profile is not affected by the galaxy background. 

%--------------------------------------------------------- King Profiles (surface brightness) and colour

\subsection{Comparison to GCs and Fornax and Virgo NSCs}
\begin{figure}
\centering
\includegraphics[width=0.49\textwidth]{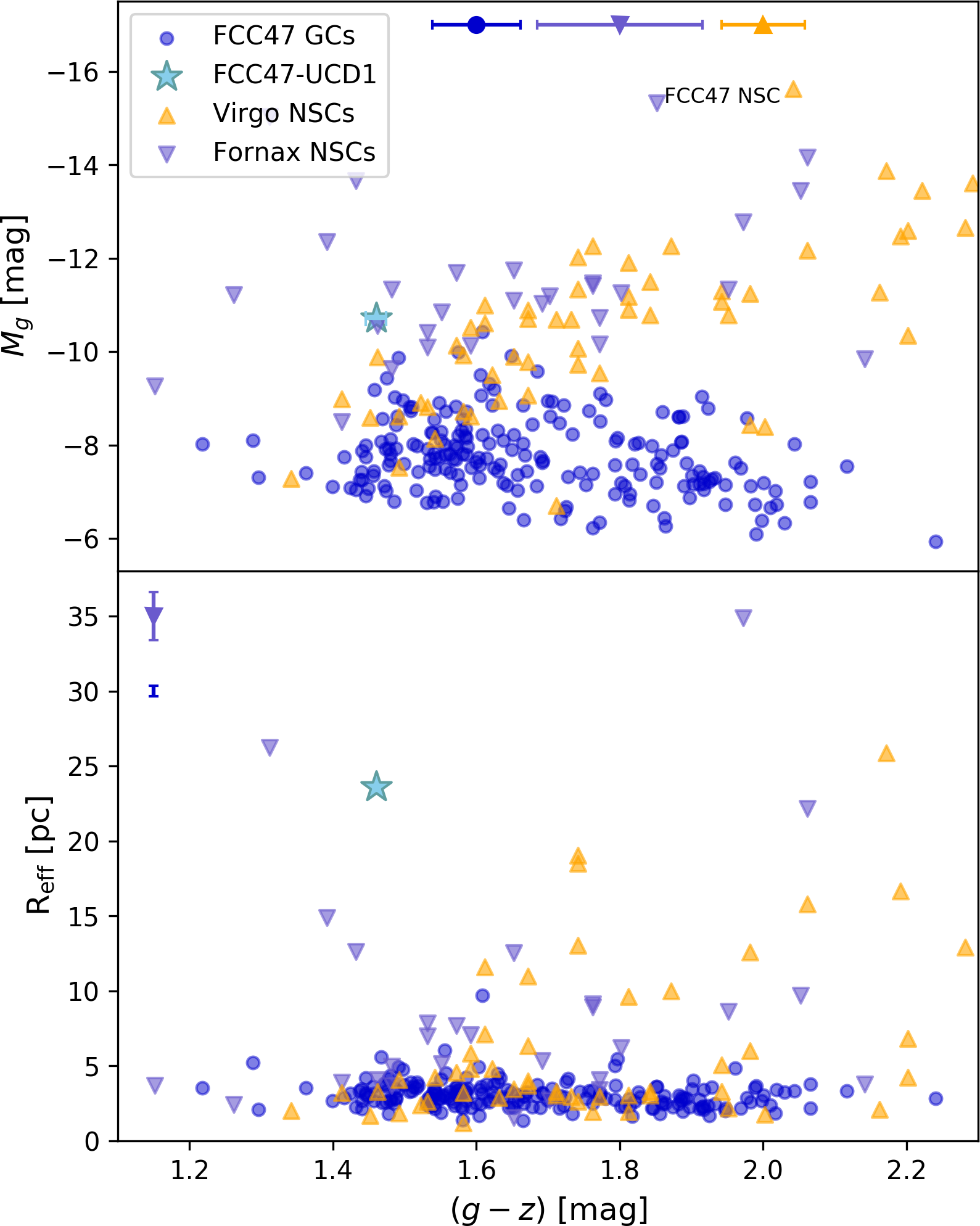}
\caption{\textit{Top}: Colour-magnitude diagram. Absolute $M_g$ magnitude vs. $(g - z)$ colours for the UCD (light blue), the GCs in FCC\,47 (blue circles, \citealt{Jordan2015}), NSCs in the Virgo cluster (orange triangles, \citealt{Cote2006}) and in the Fornax cluster (purple triangles, \citealt{Turner2012}). All magnitudes are in the Vega system. Typical colour errors are shown with the symbols on the top. The uncertainty of the UCD's colour is roughly the extend of its symbol. \textit{Bottom}: Effective radii for GCs, NSCs and the UCD. Typical uncertainties given at the left-hand side. There are no uncertainties given for the Virgo NSCs; however, we assume that they are comparable to the Fornax NSC uncertainties.}
\label{fig:UCD_CMD}
\end{figure}

We show the UCD in a colour-magnitude diagram in the upper panel of Figure \ref{fig:UCD_CMD} in comparison to the GC candidates from the ACSFCS as well as NSCs in early-type hosts in the Virgo and Fornax clusters \citep{Cote2006, Turner2012}. All magnitudes are converted to absolute Vega magnitudes using the distance moduli of $(m-M)_\text{Fornax} = 31.51$ and $(m-M)_\text{Virgo} = 31.09$ \citep{Blakeslee2009}.
The UCD is marked by the star symbol. It is about 0.5 magnitudes brighter than the next brightest GC around FCC\,47, that is also redder. The next brightest GC with a similar colour has a magnitude difference of $\sim$ 1.0 mag. It is three magnitudes fainter than the NSC of FCC\,47. With its magnitude and colour, the UCD sits at the bright end of the blue GC distribution and lies among the NSCs, especially among the fainter Fornax NSC population. It is bluer than many Fornax NSCs and brighter than Virgo NSCs of similar colour. NSCs with similar $g$-band magnitudes and colours all belong to dwarf elliptical galaxies in the Fornax cluster with typical total $B$-band magnitudes of $\sim 15.5$ mag \citep{Ferguson1989}. 
The lower panel of Figure \ref{fig:UCD_CMD} shows the effective radii of GCs, NSCs and the UCD in comparsion to the ($g$ - $z$) colour. The UCD is more than twice as large as any GC of FCC47 and larger than most of the NSCs.

\subsection{Comparison to the globular cluster luminosity function}
\begin{figure}
\centering
\includegraphics[width=0.49\textwidth]{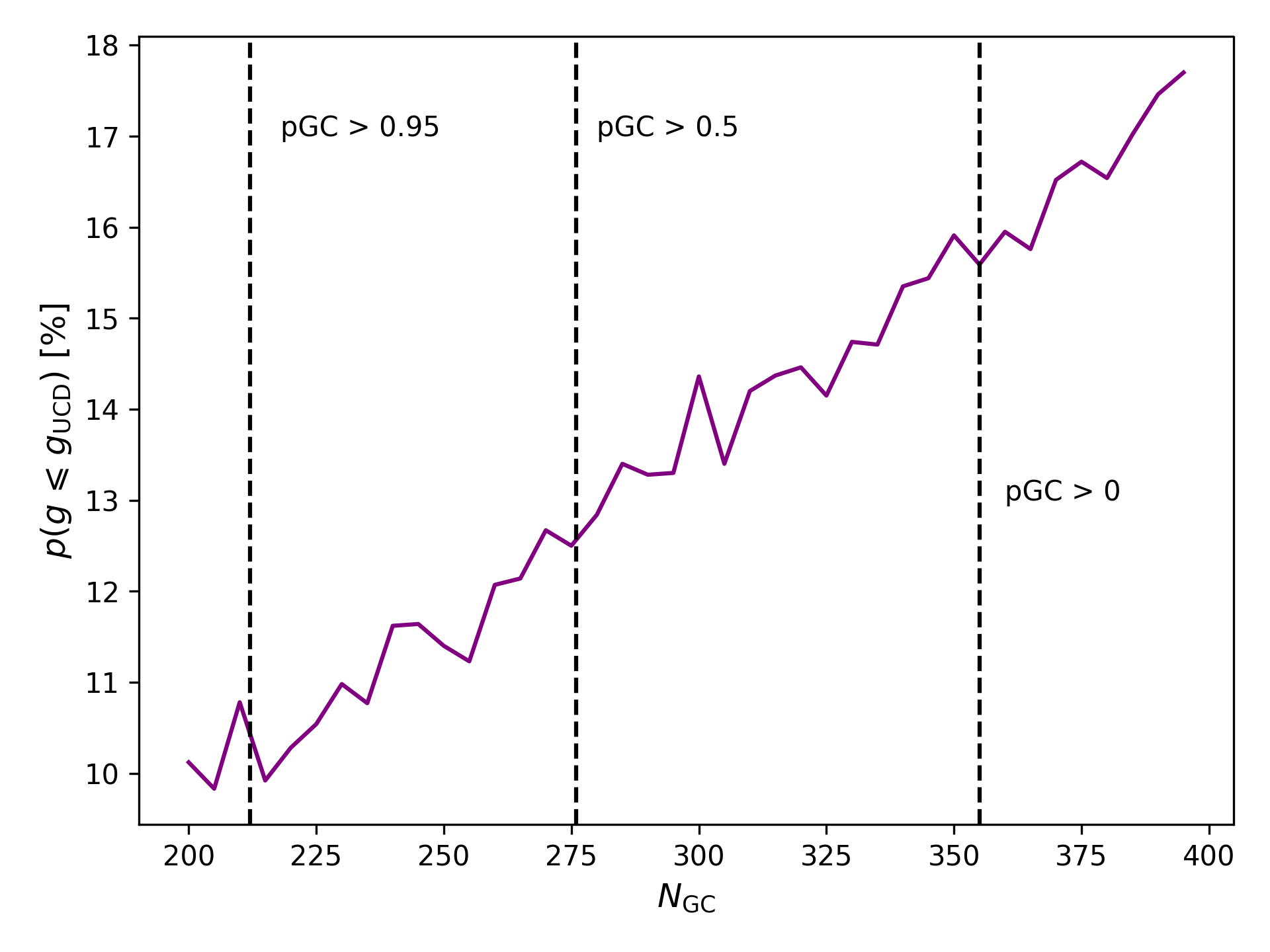}
\caption{Probability $p$ of finding at least one GC with a $g$-band magnitude less or equal to the UCD's magnitude versus the number of GCs ($N_\text{GC}$ ) that are drawn from FCC\,47's GCLF. For each $N_\text{GC}$, 10000 mock GC systems are drawn from the GCLF to compute the probability. The dotted lines indicate the number of GC candidates from the catalogue of \cite{Jordan2015} that have pGC parameter larger than 0, 0.5 and 0.95, respectively.}
\label{fig:GCLF}
\end{figure}

As richer GC systems are more likely to form more luminous GCs \citep{Hilker2009, NorrisKannappan2011}, we evaluate whether the UCDs magnitude is consistent with FCC\,47's GC luminosity function (GCLF). The GCLF of the ACSFCS galaxies was studied by \cite{Villegas2010} using GC candidates all ACSFCS galaxies that were extracted following the pipeline described in \cite{Jordan2004, Jordan2009} and were later published as catalogues by \cite{Jordan2015}. All GC candidates that have a probability parameter pGC of being a GC with pGC $>$ 0.5 are used. These are 276 candidates in FCC\,47 and their GCLF is described by a Gaussian function with a mean magnitude of $\mu$ = (23.993 $\pm$ 0.068) mag and a standard deviation of $\sigma$ = (0.988 $\pm$ 0.053) mag (in AB magnitudes) in the F475W $g$-band. We use this GCLF to randomly draw a total number of GCs ($N_\text{GC}$) to create a mock GC system of FCC\,47. This is repeated 10000 times and we then count all trails where there is at least one mock GC with a magnitude of the UCD or brighter. The resulting probability of finding at least one such GC $p$ is dependent on $N_\text{GC}$, the total number of GCs (see Figure \ref{fig:GCLF}). \cite{Harris2013} give $N_\text{GC}$ = 276 $\pm$ 100 GCs, so we assume that FCC\,47 has not more than 400 GCs, probably way less. With $N_\text{GC}$ = 276, the probability $p$ to find a GC as bright as the UCD is about 13\%. For comparison, the brightest GC ($g \sim 21$ mag) from the ACS catalogue has a probability of $\sim$ 30\%, but the probability to find at least 2 GCs with $g <$ 21 mag is $\sim$ 7\%. The UCD's magnitude is therefore consistent with FCC\,47's GCLF.
%----------------- GCLF thingy

%---------------------------Kinematics and mass estimate

\subsection{Kinematics and mass estimates}
The \textsc{pPXF} fit to the background-subtracted MUSE spectrum reveals the radial velocity of the UCD.
The UCD has a heliocentric radial velocity of $v_\text{rad} = (1509.7 \pm 4.5$) $\text{km s}^{-1}$.  The uncertainty refers to the standard deviation from the distribution in Figure \ref{fig:vs_metals_distributions} as obtained from 600 MC runs. The UCD has a relative velocity of $(65.3 \pm 4.9)$ $\text{km s}^{-1}$ compared to the systemic velocity of FCC\,47 (1444.4 $\pm$ 2.0 $\text{km s}^{-1}$, Fahrion et. al. in prep). 
The relative radial velocity of the UCD reveals that it is indeed related to FCC\,47 because the velocity dispersion of the stars in FCC\,47 is $\sim$ 80 $\text{km s}^{-1}$ at the position of the UCD (Fahrion et. al. in prep). This suggests that the escape velocity is much larger. 

As described above, we determine the upper limit of the velocity dispersion using the CaT region of the background-subtracted MUSE spectrum. From the distribution seen in Figure \ref{fig:CaT}, we find an upper limit of the velocity dispersion of $\sigma < 16.6^{+8.2}$ $\text{km s}^{-1}$. This result was obtained under the assumption that the MUSE instrumental resolution behaves as found by \cite{Guerou2017}. We also test how the instrumental resolution affects the result by repeating the fits assuming a different instrumental resolution between 2.5 and 2.9 \AA. These tests give no significantly different results, partly because of the large uncertainties.

We use the mass modelling procedure described in \cite{Hilker2007} to derive an upper limit on the dynamical mass of FCC\,47-UCD1. This approach uses the structural parameters as determined by the photometric fit of the generalized King profile and translates the 2D luminosity profile into a 3D density profile. This allows to determine the value of the 2D and 3D effective radii of the UCD based on the best-fit generalized King model parameters. These values are independent from the true velocity dispersion of the UCD. We determine the uncertainty on the effective radii by combining the parameters of the generalized King models within their uncertainties to generate the lowest and highest values possible.
Under the assumption of a certain total mass, the energy distribution function and cumulative mass profile are inferred from the density profile and are used to create a $N$-body representation of the UCD. In the last step, the observational seeing and the aperture radius are used to derive the velocity dispersion of the model as it would be observed. By comparing this value to the observed one, the input mass is iteratively adapted until we find a match to the upper limit of the velocity dispersion. We use $N$-body representations of the UCD in both filters, each time using 100\,000 particles. We list the parameters inferred from this modelling in Table \ref{tab:UCD_mass_modelling}. 

The modelling gives an upper limit of the dynamical mass of $M_\text{dyn} < 1.3^{+1.6} \times 10^7 M_\sun$ using the King model parameters in the F475W filter. The quoted uncertainty was estimated by not using the mean of the velocity dispersion distribution but the upper 1$\sigma$ deviation. Using the F850LP King model parameters, basically gives the same result. The lower limit on the dynamical mass can not be constrained with our spectrum; however, we can assume that it can not be less than the photometric mass. This upper limit is comparable to masses of other UCDs with similar effective radii (e.g. \citealt{Mieske2008}).

\begin{table}
\centering
\caption{Parameters from the mass modelling.}
\begin{threeparttable}
\begin{tabular}{c c c c}
\hline \hline
Filter &  $R_\text{eff, 2D}$ & $R_\text{eff, 3D}$ &  $M_\text{dyn}$ \\
	&  [pc]  & [pc] & [10$^7$ $M_\odot$]\\
(1)    & (2)     & (3)    & (4) \\ \hline   
F475W  ($g$) &  23.6 $\pm$ 1.4 & 31.5 $\pm$ 1.9 & $<$1.30$^{+1.60}$	\\
F850LP ($z$) &  20.6 $\pm$ 1.5 & 27.5 $\pm$ 2.0 & $<$1.32$^{+1.60}$ \\ \hline
\end{tabular}
\begin{tablenotes}
\item (1) HST filter, from the $N$-body model: (2) 2D and (3) 3D effective radius and (4) upper limit on the dynamical mass. 
\end{tablenotes}
\end{threeparttable}
\label{tab:UCD_mass_modelling}
\end{table}

\subsection{Stellar population analysis}
\label{sect:metallicity}
In addition to the kinematics, the fit with \textsc{pPXF} gives an estimate of the mean metallicity. 
While spectroscopic metallicities are in general more reliable due to the well known colour-metallicity degeneracy \citep{Yoon2011}, our MUSE spectrum has a rather low S/N and the MUSE wavelength range is not ideal to accurately constrain the age of old populations, especially at low metallicities. As mentioned above, we find that the UCD is old with a best-fitting age of $\sim$ 13 Gyr and can exclude a significant contribution from a young stellar population. From fitting the spectrum with an age-restricted SSP library, we obtain the distribution in Figure \ref{fig:vs_metals_distributions} from 600 MC runs and find a metallicity of [M/H] $=-1.12 \pm 0.10$ dex. The uncertainty refers to the standard deviation of the distribution. 

We tested the robustness of this result using different SSP templates, in particular using the so-called Padova+00 isochrones \citep{Girardi2000} instead of BaSTI isochrones or the MILES templates on a restricted wavelength range. Performing the same number of fits gives similar results within our uncertainties. Using the photometric predictions of the E-MILES SSP models and the ($g$ - $z$) colour, we find a photometric metallicity of $\sim$ -1.4 dex assuming an age of 13 Gyr. Assuming a younger age between 10 and 12 Gyr results in a better match with the spectroscopic metallicity but due to the uncertainty on the metallicity it is not possible to further constrain the age using the colour. Nonetheless, the comparison to photometric predictions confirms that FCC47-UCD1 is old (> 8 Gyr) and clearly dominated by a metal-poor population. 

We translate the metallicity into a photometric $M/L$ using the E-MILES SSP predictions. Unfortunately, the SSP predictions for $M/L$ do not cover the F850LP filter, but we can use the $M/L$ prediction for the F475W filter. We interpolate the given grid of SSP models to draw 5000 values of the $M/L$ ratio using a Gaussian distribution of the metallicity with a mean of $-$1.12 dex and standard deviation of 0.10 dex. As before, we assume an age of 13 Gyr. We find $M/L_\text{SSP, g} = 2.42 \, ^{+0.08}_{-0.10} \frac{M_\sun}{L_\sun}$ and with a total luminosity of the UCD of $L_\text{UCD} = 2.01 \times 10^6 L_\sun$ based on it's $g$-band magnitude and a distance of 18.3 Mpc, this translates to a photometric stellar mass of $M^\text{g}_\ast = 4.87 \, ^{+0.21}_{-0.16} \, \times 10^6 \, M_\odot.$
This is roughly half of the upper limit we derived for the dynamical mass. While $M/L_\text{SSP}$ = 2.42, we find an upper limit of $M/L_\text{dyn} < 6.47$.

%----------------------------------------------------------------- 

%----------------------------------------------------------------- DISCUSSION 

\section{Discussion}
\label{sect:discussion}
\subsection{UCD properties}
\label{sect:ucd_properties_discussion}
In the following, we will discuss the properties of FCC\,47-UCD1 with respect to both UCD formation scenarios: The formation as genuine star cluster and as the remnant NSC of a stripped galaxy.

In previous studies, the light profile of many massive UCDs ($M > 10^7 M_\sun$) have been found to be best represented by a dense central component together with a more extended diffuse halo \citep{Evstigneeva2008, Strader2013, Ahn2018, Afanasiev2018}. With respect to simulations by \cite{Pfeffer2013}, this two component structure is in agreement with the stripped nucleus formation scenario, where the central component is interpreted as the remnant NSC of the UCD progenitor while the diffuse component is interpreted as the remains of the stripped galaxy. In the case of the FCC\,47-UCD1, we find that the structure is best represented by a single generalized King model. We do not detect a second component that could represent a stellar envelope. This is consistent with FCC\,47-UCD1 being a high-mass star cluster with no extended envelope and a large effective radius of 24 pc, but a small envelope might also be hidden by our resolution limit and might cause the large measured effective radius. \cite{Pfeffer2013} explored the formation of UCDs via tidal stripping of nucleated dwarf galaxies with simulations and found that extended envelopes fall below our surface brightness limit of 25 mag arcsec$^{-2}$ already 2 Gyr after the stripping.

FCC\,47-UCD1 has a $g$-band magnitude of $g = 20.76$ mag and is therefore $\sim$ 0.5 magnitudes brighter than the brightest GC from the ACSFCS sample \citep{Jordan2015}. As our consideration of the GCLF shows, the probability to find a GC with this magnitude from FCC\,47's GCLF is $\sim$ 13 \%. The UCD magnitude is therefore still consistent with the GCLF. With this magnitude, FCC\,47-UCD1 is fainter than the previously confirmed NSC-type GCs.

\begin{figure}
\centering
\includegraphics[width=0.49\textwidth]{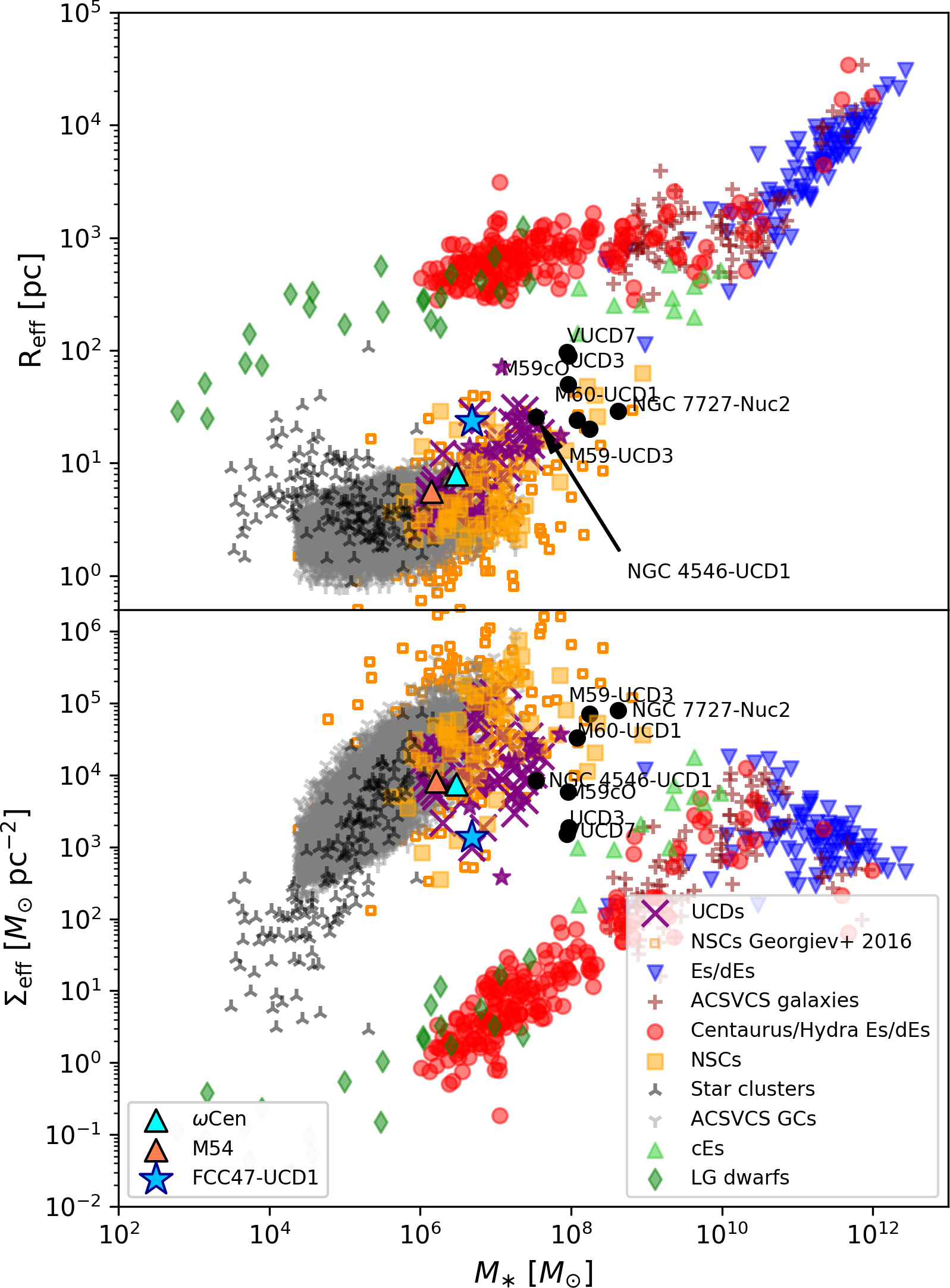}
\caption{\textit{Top}: Effective radii of stellar systems versus stellar mass. \textit{Bottom}: Effective surface mass density ($\Sigma_\text{eff} = M_\ast/2 \pi R_\text{eff}$) versus stellar mass. The data was mostly collected and presented in \cite{Misgeld2011}. We include a few more UCDs from \cite{NorrisKannappan2011, Liu2015, Sandoval2015, Schweizer2018} and NSCs from \cite{Georgiev2016}. Known NSC-type UCDs are marked by the solid black dots as well as M54 and $\omega$Centauri. FCC\,47-UCD1 is given by the blue star.}
\label{fig:mass_density}
\end{figure}

The core radius obtained from the generalized King models of $\sim$ 11 pc is larger than what was found for many other UCDs in Fornax and Virgo studied by \cite{Mieske2008}, but we find comparable values for the $\alpha$ parameter that gives the slope of the density profile and the effective radius of 24 pc is also similar to other UCDs. For a more thorough comparison, we put the UCD in mass-size and mass-surface mass density plots in Figure \ref{fig:mass_density} similar to those presented, e.g., in \cite{Misgeld2011, Norris2014, Sandoval2015}. The collected data is mostly taken from \cite{Misgeld2011} and references therein, but we also include a few UCDs from \cite{Liu2015},  \cite{NorrisKannappan2011} and \cite{Schweizer2018} with available mass measurements. Known NSC-type UCDs are marked as well as M54 and $\omega$Cen \citep{Harris1996, Baumgardt2017}. The latter are officially classified as GCs of the Milky Way, but most likely are the stripped nuclei of accreted dwarf galaxies (e.g. \citealt{Zinnecker1988, Ibata1997, HilkerRichtler2000, King2012, Bellazzini2008}). FCC\,47-UCD1 is larger than similar bright GCs and larger than most of the shown NSCs and UCDs. Its size is comparable to some diffuse star clusters, but these have much lower masses (e.g. \citealt{Liu2016}). The comparison to known NSC-type UCDs shows that these are more massive, bigger and denser than FCC\,47-UCD1.

As UCDs are considered to be unrelaxed systems \citep{Fellhauer2002,NorrisKannappan2011, Voggel2018}, a colour gradient translates directly into a population gradient (instead of being an effect of mass segregation), where a slightly younger population is found at the center while the outskirts are dominated by a older, metal-poor population. Although we cannot find a significant colour gradient, we find the UCD to appear larger in the blue than the red filter by $\sim$ 10\%, corresponding to a more extended metal-poor population.
As FCC\,47-UCD1 is unresolved in the MUSE data, we cannot study a possible gradient spectroscopically. Our \textsc{pPXF} fit gives no indication of multiple populations, possibly because the S/N is insufficient to disentangle multiple populations with this small colour difference. 

The MUSE spectrum gives a mean metallicity of \mbox{[M/H] = $-1.12$ dex}, which is metal-poorer than what was found for most massive UCDs (e.g. \citealt{Mieske2013, Janz2016, Seth2014, Ahn2018}) In particular, FCC\,47-UCD1 is clearly more metal poor than the confirmed NSC-type UCDs that are typically characterised by elevated metallicites (e.g. \citealt{Janz2016}). However, there are other UCDs with similarly low metallicities, such as NGC 3923-UCD2 and UCD3 or S999 \citep{NorrisKannappan2011, Janz2016}. The metallicity is comparable to some of the GCs found in FCC\,47 as will be presented in Fahrion et. al. (in prep.) while other GCs have a higher metallicity.  The low metallicity and old age of the UCD is consistent with it being a genuine star cluster similar to the blue GCs found in FCC\,47; however, many NSCs of dwarf galaxies are known to have a similar low metallicity \citep{Paudel2011, Sandoval2015}.

The upper limit of the ratio of dynamical-to-photometric $M/L$ $\Gamma < 2.68$ is above the average value of 1.7 \citep{Mieske2013}, but well within the scatter and the true value could be well below. If higher spectral resolution observations would find an elevated dynamical $M/L$, this could indicate the presence of a central BH as found in other, more massive UCDs \citep{Mieske2013, Seth2014}. Alternatively, it could be attributed to IMF variations \citep {MieskeKroupa2008, Mieske2013} or tidal stripping \citep{Forbes2014, Pfeffer2013}. However, the low mass and apparent magnitude of FCC\,47-UCD1 limit the deep spectroscopic and/or photometric observations that would be needed to detect these features.

The presented photometric, kinematic and structural properties of FCC\,47 do not allow to unambiguously pinpoint its origin. The magnitude, colour, metallicity and the absence of an extended envelope are consistent with it being a very massive star cluster; however, the stripped NSC of a dwarf galaxy can show similar properties (see e.g. \citealt{Pfeffer2013}). The very large effective radius of $\sim$ 24 pc makes it a complete outlier from the rest of FCC\,47's GCs. This is perhaps the most convincing argument that this object is different from typical GCs, making FCC\,47-UCD1 a candidate of the stripped nucleus of a dwarf galaxy, similar to the ones we find in the MW (e.g. $\omega$Cen and M54).

\subsection{Implications for the stripped nucleus scenario}
Under the assumption that FCC\,47-UCD1 is the remnant nucleus of a stripped dwarf galaxy, in this section we explore the properties of its hypothetical progenitor galaxy.

\subsubsection{Comparison to Fornax NSCs}
As the colour magnitude diagram reveals (Figure \ref{fig:UCD_CMD}), the magnitude and colour of FCC\,47-UCD1 is comparable to several NSCs of galaxies in the Fornax clusters. We find that the UCD has almost the same colour and magnitude as the NSC of FCC19, a nucleated dwarf elliptical galaxy with an $B$-band magnitude of 15.2 \citep{Ferguson1989}. Assuming a $M/L_\text{B}\sim2$ as typical for such a galaxy \citep[e.g.][]{BelldeJong2001}, FCC19 probably has a mass of a few $10^{8} M_\sun$. We find $B$-band magnitudes for the galaxies FCC100 and FCC202, two other nucleated dwarf galaxies that both have NSCs with similar brightness as FCC\,47-UCD1 but at a redder colour. While there is no mass estimate available for FCC\,100, \cite{Eigenthaler2018} report a stellar mass of $M_\ast = 1.9 \times 10^9 M_\sun$ for FCC\,202. Under the assumption that FCC\,47-UCD1 would be the stripped nucleus of a galaxy, it is likely that its progenitor had a similar mass as FCC\,19 has today. However, we can not determine if the progenitor was an ellipcital dwarf galaxy like the dwarfs we see in the Fornax cluster today or if it rather had a spiral/irregular morphology. Some dwarf irregulars are known to host massive old clusters in their nuclear regions that are separated from the normal GCs by a magnitude gap of $\sim$ 0.5 mag \citep{Georgiev2009b}.

\subsubsection{NSC-galaxy mass relation}
It has long been established that the properties of nuclear regions of galaxies, such as the central SMBH or the NSC, scale with properties of the galaxy itself \citep{Magorrian1998, Ferrarese2006, GrahamDriver2007, KormendyHo2013}. \cite{Georgiev2016} studied the relation between stellar mass of the NSC and host stellar mass for both early- and late-type galaxies (ETGs/LTGs) with masses $> 10^8 M_\sun$. In case FCC\,47-UCD1 indeed is of NSC-origin, we can use these relations to predict the mass of the UCD progenitor if we assume that its stellar body has been stripped entirely without altering the nucleus. Assuming an early-type progenitor, we find a progenitor mass of $M_{\rm \star, prog, ETG} = (3.1 \pm 0.3) \times 10^9 M_\sun$, and $M_{\rm \star, prog, LTG} = (5.9 \pm 0.6) \times 10^9 M_\sun$ assuming a late-type progenitor. This is around 10 times larger than the typical mass of a dwarf galaxy and probably overestimates the mass of a hypothetical progenitor of FCC\,47-UCD1 as the used relations were derived using giant galaxies rather than dwarfs. 

\cite{Spengler2017} did a similar study using NSCs in the Virgo cluster to extend the relation of \cite{Georgiev2016} to lower NSC and host masses. The authors give no analytical expression for their relation, but found that the relation between NSC mass and host mass flattens for host masses below $10^8 M_\sun$. 
Their relation shows a large scatter that can comprise up to two orders of magnitude for a given nucleus mass, especially for nuclei with masses $< 10^6 M_\sun$. For the stellar mass of the UCD, a stellar mass of the progenitor between $10^7$ and $10^9 M_\sun$ is within the scatter. 

Using nucleated dwarf galaxies observed in the Next Generation Fornax Survey, \cite{Ordenes2018} extended the relation between nucleus mass and host mass down to host masses of $10^6 M_\sun$. The authors also find a flattening of the canonical relation towards lower masses. Using this relation and the photometric mass estimate of the UCD, the estimated mass of a possible progenitor galaxy is $M_{\ast, \rm{prog}}$ = ($7.2 \pm 0.4) \times 10^8 M_\sun$. 
From the presented NSC-galaxy mass relations, the last is likely the most applicable one for estimating a possible progenitor galaxy of FCC\,47-UCD1 because it was derived using dwarf galaxies while the other can still give a reasonable order of magnitude estimate.

\subsubsection{Mass-metallicity relation} %----------------------------
It is well established that the average metal content of a galaxy is correlated with its stellar \citep{Lequeux1979, Tremonti2004, Gallazzi2005, Mendel2009}. 
Studies in the Fornax and Virgo clusters have shown that UCDs typically have high metallicities for their luminosities and lie above the canonical metallicity-luminosity trend for elliptical galaxies \citep{Chilingarian2011, Francis2012}. Instead, they seem to follow the metallicity-luminosity trend of dwarf galaxies.
For dwarf galaxies with stellar masses below $\sim 10^{10}$ $M_\sun$, \cite{Kirby2013} have found a tight relationship between stellar mass and iron metallicity:
\begin{equation}
    [\text{Fe/H}] = (-1.69 \pm 0.04) + (0.30 \pm 0.03)\, \text{log}\left(\frac{M_\ast}{10^6 M_\sun}\right)
   \label{eq:metallicity_relation}
\end{equation}
As the used E-MILES SSP models give [M/H] values instead of direct measurements of the iron abundances, we test the difference between the two by fitting the MUSE spectrum again with scaled solar MILES models that have [M/H] = [Fe/H] with no alpha-enhancement. The original MILES SSP models have a restricted wavelength range between 3500 and 7500\AA, so redder wavelengths are not used in this comparison. As we find no difference in the resulting metallicity, we assume that the iron metallicity of the UCD is the same as the total metallicity within our uncertainties. 

We use the relation in Equation \ref{eq:metallicity_relation} to estimate the stellar mass of a potential UCD progenitor under the assumption that the measured UCD metallicity represents an average of the progenitor galaxy. Usually, the highest metallicities are found at the center of galaxies \citep{Couture1988, Koleva2011}, so that the UCD's metallicity would give an upper limit on the stellar mass of a possible progenitor. However, this depends on the time of the stripping as it is possible that the progenitor galaxy today might have a higher metallicity from continuing star formation if it were not stripped. This is supported by the finding that UCDs typically have a lower metallicity than NSCs of a similar mass \citep{Paudel2011, Spengler2017}, because they are dominated by star cluster-type UCDs. Confirmed NSC-type UCDs have metallicities comparable to massive NSCs. We therefore assume that the UCD metallicity is a lower limit of what the UCD progenitor would have today without stripping and consequently the resulting mass gives a lower limit.

Using the spectroscopic [M/H] = $-$1.12 $\pm$ 0.10 dex, the estimated stellar mass of a hypothetical UCD progenitor is \mbox{$M_{\rm \ast, prog}= (7.94 \pm 0.24) \times 10^7 M_\sun.$}

Under the assumption that FCC\,47-UCD1 is a stripped NSC, the presented scaling relations imply that its potential progenitor was a dwarf galaxy similar to those found today in the Fornax cluster with a mass of a few $\sim 10^8 M_\sun$. In this case, the comparison to the photometric mass of \mbox{$M_{\rm SSP} = 4.87 \times 10^6 M_\sun$}, this implies that the UCD progenitor lost $> 95\%$ of its mass in a $\sim$ 1:100 merger with FCC\,47.

Such a dwarf galaxy might have brought a few GCs with it that should now be in the halo of FCC\,47 among the blue GC population. However, identifying those in FCC\,47's rich GC system is challenging.

\subsection{Implications for star cluster origin}
The low metallicity, old age and magnitude of FCC\,47-UCD1 are consistent with it being a very massive extended star cluster. In the MW, no GCs with this mass and size are known, but they have been observed in other galaxies (e.g. in M87 \citealt{Strader2011}) and \cite{Murray2009} proposed that very massive GCs might follow a mass-size relation leading to an extended size. In this case, FCC\,47-UCD1 would mark the high-mass (high luminosity) end of the blue GC population. Here, we want to explore the implications in the case FCC\,47-UCD1 is indeed a genuine star cluster.

\subsubsection{Dynamical friction}
UCDs and massive star clusters that are close to their host's centre should spiral in-wards due to dynamical friction.
The timescale of inspiral of a satellite on a circular orbit in a singular isothermal sphere with velocity dispersion $\sigma_{\rm{sat}}$ from an initial radius $r_i$ to the centre of a galaxy with characteristic velocity dispersion $\sigma_\text{gal}$, is given by \citep{BinneyTremaine1987}:
\begin{equation}
t_\text{fric} = \frac{2.7\, \text{Gyr}}{\text{ln} \Lambda} \frac{r_i}{30 \, \text{kpc}} \left(\frac{\sigma_\text{gal}}{300 \, \text{km s}^{-1}}\right)^2 \left(\frac{\sigma_{\rm{sat}}}{100 \, \text{km s}^{-1}}\right)^{-3},
\end{equation}
where $\Lambda$ is the Coulomb logarithm. $\Lambda = 2^{2/3} \sigma_\text{gal}/\sigma_s$ under the assumption that the satellite has a significantly smaller dispersion than the galaxy. For FCC\,47, $\sigma_\text{gal} \approx 80$ $\text{km s}^{-1}$, as we determine from the velocity dispersion map presented in Fahrion et. al. in prep. 

%For an initial radius of 50 kpc, the inspiral time is already above 9 Gyr. 
If we assume that FCC\,47-UCD1 is on such a circular orbit around the galaxy with its 3D separation given by the projected distance of 1.1 kpc, the inspiral timescale to the centre is less than 0.5 Gyr. 
The short dynamical timescale allows different interpretations. Firstly, if the UCD is of GC-origin a near-circular orbit is plausible, implying that the UCD will soon merge into the NSC of FCC\,47. In this case, it must have formed much further out in the outskirts of FCC\,47 to still be observed today. Choosing an inital radius of $\sim$ 50 kpc, the in-spiral time increases to $>9$ Gyr. Such a large initial distance can explain why FCC\,47-UCD1 is still observable today, but the formation of such a massive star cluster might be difficult at large galactocentric radii due to the low gas density, even at early times (e.g. \citealt{Pfeffer2018}). 
Following the argumentation of \cite{NorrisKannappan2011} that have compared the dynamical timescale of GC-type and NSC-type UCDs, long-lived (old) massive objects such as FCC\,47-UCD1 are not expected to have these short dynamical timescales on average if they are of GC-origin because only a few should be caught by chance right before their merger into the galaxy centre.

The dynamical friction timescale as presented here assumes a circular orbit in an isothermal sphere; however, the assumption of a isothermal sphere might not be valid because the dynamical orbit-based Schwarzschild model of FCC\,47 as presented in Fahrion et. al. (in prep.) shows that FCC\,47 is a triaxial galaxy. This model also shows that the circular velocity at the UCD's position is around $\sim$ 100 km\,s$^{-1}$, whereas the UCD has a relative velocity of $\sim$ 65 km s$^{-1}$. This could indicate that the projected distance gives an underestimation of the true 3D separation under the assumption of a circular orbit, which in turn would also increase the dynamical friction timescale. The assumption of a circular orbit might also not apply in case FCC\,47-UCD1 is of ex-situ origin. If it was located in an accreted dwarf galaxy, it is more likely that it came in on a radial orbit that has on average a much larger dynamical friction timescale.

\subsubsection{Ex-situ origin}
In the context of the hierarchical assembly of galaxies, the blue GC subpopulation is often interpeted as having formed in low-mass dwarf galaxies and then being accreted during merger events \citep{Cote1998, Hilker1999, Lotz2004, Peng2006, Cote2006, Georgiev2008}. Indications of such an ex-situ origin is often given by stark contrasts in chemical properties between the host galaxy and the GC/UCD \citep{NorrisKannappan2011} and the red GC system \citep{Cote2006}. The stellar body of FCC\,47 at the UCD's location is dominated by a old, gas-free population that is much more metal-rich ([M/H] $\sim -0.4$ dex) than the UCD. As will be presented in an accompanying paper (Fahrion et. al. in prep), FCC\,47 has a bi-modal GC population that contains both GCs with metallicities similar to the UCD and more metal-rich GCs. Together with the short dynamical friction timescale, the low metallicity and blue colour are indications of an ex-situ origin of the UCD, irrespectively of its GC or NSC-nature.

Under the assumption that FCC\,47-UCD1 was accreted during the assembly of the host galaxy, it is likely that it has been the most massive star cluster of its original host galaxy. As \cite{Hilker2009} showed, the absolute magnitude of the brightest GC/UCD in a galaxy scales with the host luminosity and total number of GCs. FCC\,47-UCD1 falls among this relation as the brightest GC of FCC\,47's rich GC system, but if it came from an accreted satellite, this dwarf galaxy must have had a smaller number of GCs than FCC\,47 has today and thus it is difficult to reconcile the brightness of the UCD with a much smaller GC system of such a hypothetical progenitor following the relation described in \cite{Hilker2009}. However, this discrepancy is naturally solved if FCC\,47-UCD1 was not a nominal GC of its progenitor host, but instead the NSC; similar e.g. to M54 which is considered to be the metal-poor NSC of the Sagittarius dwarf galaxy \citep{Ibata1997, Bellazzini2008}. At these low galaxy masses, colours and metallicities of dwarf nuclei are no longer distinguishable from blue GCs or UCDs \citep{NorrisKannappan2011, Sandoval2015, Spengler2017}. Using HST/ACS, \cite{Georgiev2009a} studied the GC populations of late-type dwarf galaxies and found many massive star clusters with magnitudes $M_V < -$10 mag. These massive GCs are found at small projected distances in the nuclear regions of these dwarf galaxies and show similar properties concerning the mass, size and metallicity as M54 or $\omega$Cen. In the case that FCC\,47-UCD1 was accreted, it is possible that it has been the nuclear cluster of such a metal-poor dwarf satellite galaxy.

\subsection{On the UCD projected distance}
\begin{figure*}
    \centering
    \includegraphics[width = 0.98\textwidth]{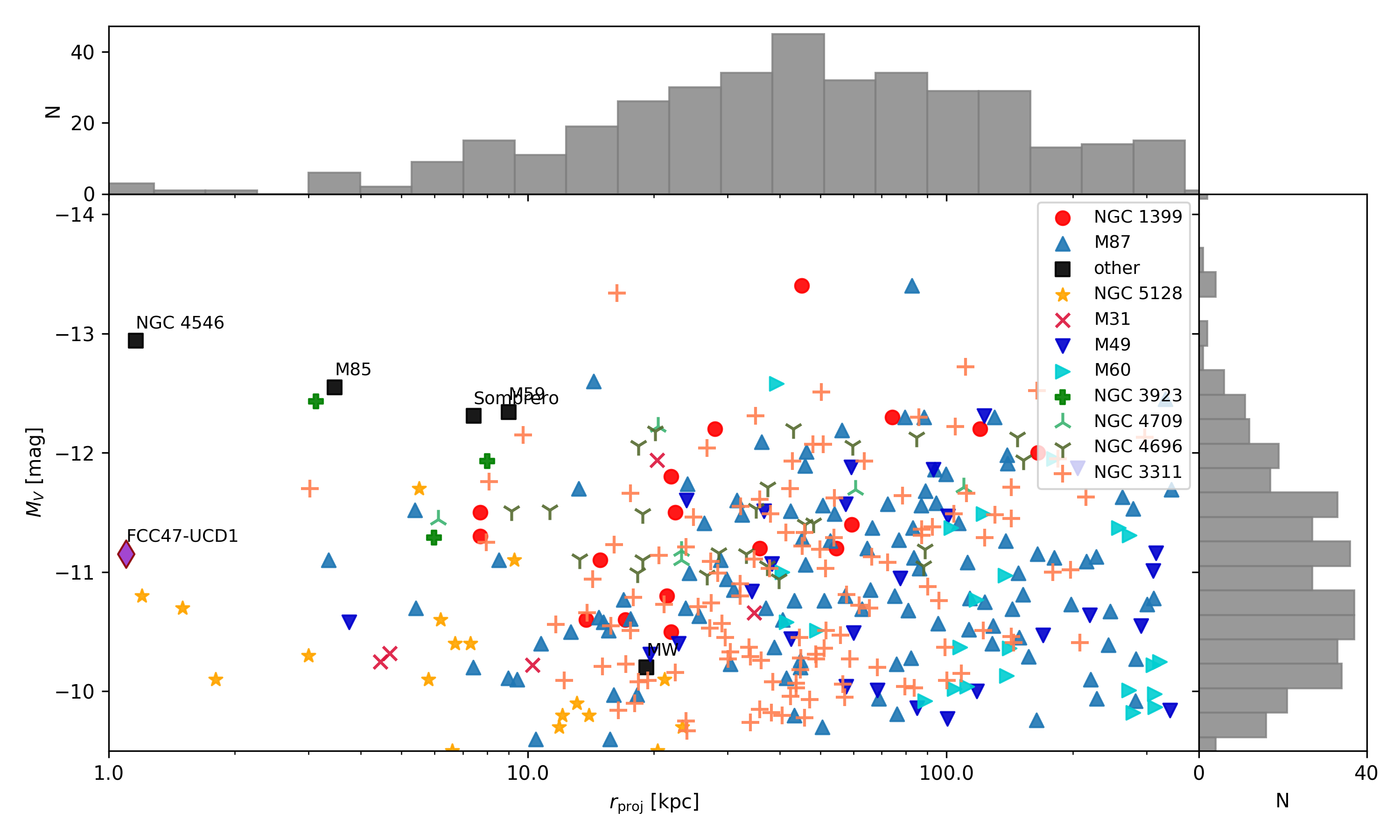}
    \caption{Projected distances of UCDs to their assumed host galaxies vs their magnitudes. We show the UCDs from several sources including \cite{Rejkuba2007, Mieske2008,  Brodie2011, NorrisKannappan2011, Liu2015, Misgeld2011b, Galleti2004}. We show the sources with absolute magnitudes between -9.5 $> M_V >$ -14.0 mag, the magnitude range for typical UCDs. Details on how the data were compiled can be found in the Appendix (Section \ref{sect:UCD_catalog})}.
    \label{fig:UCD_catalog}
\end{figure*}

The UCD in FCC\,47 is special in two ways: it is found with a very close projected distance to the centre of FCC\,47 and it is bound to a rather low-mass host. Nowadays, $>300$ UCDs are known. Most of them were found using large-scale imaging surveys combined with spectroscopic follow-up observations of the centres of galaxy clusters such as Virgo \citep{Brodie2011, Liu2015}, Fornax \citep{Hasegan2005, Mieske2004} or even the Coma cluster \citep{Price2009}. Only a few were found around group or isolated galaxies, such as Centaurus A \citep{Rejkuba2007} or the Sombrero galaxy \citep{Hau2009, NorrisKannappan2011}.
Figure \ref{fig:UCD_catalog} shows a comprehensive, but probably not complete collection of the currently known UCDs with their absolute $V$-band magnitudes versus projected distances to the assumed hosts. We describe in Appendix \ref{sect:UCD_catalog} how this sample was compiled. 
This figure reveals that there are only a few UCDs with distances $<5$ kpc from their host. This is most likely dominated by a selection effect because searches for UCDs are easier in the outskirts of galaxies where the local surface brightness is low.

\cite{NorrisKannappan2011} reported on the discovery of a massive UCD close to the centre of the early-type S0 galaxy NGC\,4546 using archival HST data in addition to SAURON IFU data. Although NGC\,4546 is an isolated galaxy, the two systems are quite similar. Both FCC\,47 and NGC\,4546 are early-type galaxies with masses of $\sim 2 \times 10^{10} M_\sun$ \citep{NorrisKannappan2011} and their UCDs were found at close projected distances of $\sim$ 1.1 kpc. However; with a total stellar mass of $\sim 3 \times 10^7 M_\sun$ NGC\,4546-UCD1 is much more massive than FCC\,47-UCD1, but has a similar size ($R_\text{eff} \sim 25$ pc). This is also reflected in its magnitude ($M_V \sim -13$ mag) and supersolar metallicity ([M/H] $\sim$ 0.2 dex). At a distance of $\sim$ 13 Mpc, deep spectroscopic observations with Gemini/GMOS were possible that allowed to discover an extended star formation history \cite{Norris2015}, although no central SMBH could be detected. The large magnitude gap to the GC system in addition to the extended SFH confirmed this object to be the remnant nucleus of a stripped dwarf galaxy.

The location of FCC\,47-UCD1 in close projected distance to FCC\,47's NSC, is also reminiscent of the two nucleus set-up found in the massive merger galaxy NGC\,7727 \citep{Schweizer2018}. While the primary nucleus seems to fit well into the central luminosity and colour profile, the authors argue that the second nucleus is the stripped remnant of a former companion galaxy and therefore gives direct evidence for the formation of a UCD through stripping. The projected separation between the two nuclei (r$_\text{proj} = 0.48$ pc) is even smaller than in FCC\,47; however, there are obvious differences between the two systems: NGC\,7727 is much more massive \mbox{($M_\ast \sim 1.4 \times 10^{11} M_\sun$)} than FCC\,47 and has a complex structure due to mergers. With a $V$-band magnitude of $M_V$ = $-$15.5 mag, the second nucleus of NGC\,7727 is much brighter than typical massive UCDs and might therefore reflect a young UCD in the formation process. Although the second nucleus in NGC\,7727 is much more massive than FCC\,47-UCD1, it is possible that the NGC\,7727 systems resembles what has happened to FCC\,47 and its UCD.

%----------------------------------------------------------------- UCD CATALOG 

\subsubsection{Implications for other galaxies}
By construction, the searches for UCDs are biased towards large distances from the host galaxies, where the galaxy background is faint. FCC\,47-UCD1 is the exception that confirms the rule as it is visible in both the HST data from the ACSFCS and the VST data used in the FDS, but was so far overlooked because of its small separation to FCC\,47. One might ask whether the typically large UCD-host-separations are purely due to selection effects in the search, or whether this is rooted in their formation and evolution.

Studying the nature of the, on average, large separation of UCDs to their host galaxies, as compared to GCs, could also help to further constrain their nature as pure star cluster objects or remnant nuclei of dwarf galaxies. As more massive galaxies can form more massive star clusters \citep{Schulz2016}, star-cluster UCDs should be found in these high mass galaxies, but at large separations because, at smaller separations, they would quickly merge into the centre because their native orbits would lead to small dynamical friction timescales. On the other hand, the remnant nuclei of stripped dwarf galaxies could also be found at smaller separations if the dwarf galaxy had a radial (or polar) orbit with respect to the host galaxy and is now near its pericentre distance or on an circularized orbit. Studying the presence of UCDs (and star clusters) with close projected distance to their host's centre can also help to constrain the underlying density profile of the host because cored profiles can stall dynamical friction significantly \citep{Goerdt2006, Read2006, Cole2012}.

The fact that the UCD in FCC\,47 is bound to a rather low mass host is also of interest. Typically, UCDs have been found in the centres of galaxy clusters and are by default associated to the most massive galaxy in the cluster (e.g. \citealt{Hilker2007, Brodie2011, Liu2015}). Only a few UCDs have been found that are associated with a more isolated galaxy such as NGC\,4546 or the Sombrero galaxy \citep{NorrisKannappan2011}.

While it was time consuming in the past to first find a UCD candidate in large-scale photometric data and then to confirm its nature with a spectrum, integral field spectroscopy with instruments like MUSE could potentially reveal a large number of UCDs even in the high surface brightness regions of galaxies as it was the case for FCC\,47-UCD1. This could give insight into the radial distribution of UCDs and into possible correlations with host properties such as mass and morphology. 

%---------------------------------------------------------------------
\section{Conclusions}
\label{sect:conclusion}
In this paper we report on the discovery of a UCD near the centre of the early-type galaxy FCC\,47 using AO-supported MUSE WFM SV and archival HST data. We present a photometric analysis of the structure of this UCD and an analysis of its integrated MUSE spectrum. From combining those data and results we estimate the dynamical and stellar population mass of the UCD. 
We summarize our findings and their implications as follows:

\begin{itemize}
\item{The UCD lies at a very close projected distance (1.1 kpc) from the centre of FCC\,47. It is the second known UCD at such close projected distance to its host centre. This might be partly due to observational biases in searches of UCDs around high surface-brightness galaxies or due to very short dynamical friction timescales. Assuming a circular orbit in a isothermal sphere, FCC\,47-UCD1 should fall into the NSC within the next 0.5 Gyrs. This might imply that similarly massive and close UCDs might have already merged with their host galaxy nucleus, although a radial orbit significantly decreases the timescale and FCC\,47 seems to be an triaxial galaxy.}

\item{The UCD structure is fitted in 2D with \textsc{imfit}, finding that the best representation of the data is achieved with a single generalized King profile with an effective radius of $\sim$ 24 pc. (See Table \ref{tab:UCD_params}). With this size, the UCD is more than twice as large as any GC in FCC\,47.}
      
\item{The UCD has a colour of $(g - z) = 1.46$ and is not distinct from FCC\,47's blue GC population. Its absolute magnitude of $M_g = -10.55$ mag is consistent with the high-luminosity end of FCC\,47's GC population, but is also comparable to magnitudes of known NSCs of dwarf elliptical galaxies in the Fornax cluster}.

\item{From the MUSE spectrum, we find a relative velocity of $\sim$ 65 $\text{km s}^{-1}$ confirming membership to the FCC\,47 system and an upper limit of the velocity dispersion of $\sigma<17$ $\text{km s}^{-1}$. The inferred limit on the dynamical mass is \mbox{$M_\text{dyn} < 1.3^{+1.6} \times 10^7 M_\sun$} and we find a upper limit on the dynamical mass-to-light ratio of $M/L_\text{dyn} < 6.47$.}

\item{We find a spectroscopic metallicity of [M/H] = $-$1.12$\pm$ 0.10 dex and a generally old age ($> 8$ Gyr). Our data does not give any indication of multiple or young populations. We can exclude a significant contribution to the spectrum from a young stellar population.}

\item{From the spectroscopic metallicity, we determine the stellar population mass of $M_\text{SSP} = 4.87 \times 10^6 M_\sun$. This mass can be considered the lower limit of the total mass.}

\item{The low metallicity, old age and magnitude make this UCD consistent with being the most massive, blue GC in FCC\,47's large GC system. The blue colour, low metallicity compared to FCC\,47 stellar body and red GC population and dynamical arguments suggest an ex-situ origin. Although we cannot unambiguously confirm its nature, FCC\,47-UCD1 could be a candidate of the remnant NSC of a stripped metal-poor dwarf galaxy.}

\item{In the case the UCD is the remnant NSC of a stripped dwarf galaxy, we use known scaling relations, to estimate that its hypothetical progenitor could have been a dwarf galaxy with a mass of a few 10$^8  M_\sun$. This would imply that the progenitor has lost at least 95\% of its initial mass in an minor merger event with a mass ratio of $\sim$ 1:100.}
\end{itemize}

There are open questions that would need an answer to further characterize the UCD progenitor and the UCD itself. 
Additional data is needed to determine the true velocity dispersion of the UCD and in turn determine its true dynamical mass. For this, a spectrum with high S/N and high spectral resolution is needed. In case the elevated dynamical mass-to-light ratio remains, its origin should be investigated. While variations in the IMF might be discovered with a high S/N spectrum in the optical or infra-red wavelength range, a detection of a central SMBH would need AO-supported IFU observations. Unfortunately, at the distance of the Fornax cluster, this UCD is probably too faint to be observed with current high angular resolution IFU instruments.

The serendipitous discovery of this UCD shows how powerful the MUSE instrument is with its wide field of view and high spatial resolution. Now, with the commissioning of the GALACSI AO system, this power to discover new objects has further increased because with AO one can not only achieve superb resolution and in turn detect fainter and more distant UCDs, but the system is also capable to acquire good quality data under less than ideal conditions. 

\begin{acknowledgements}
We want to thank the anonymous referee for comments that helped to improve this manuscript.
This work is based on observations collected at the European Organization for Astronomical Research in the Southern Hemisphere under ESO programme 60.A-9192.
KF is grateful to Dimitri Gadotti and Enrica Iodice for their contributions to the original MUSE SV proposal and especially to Lodovico Coccato who kindly did the data reduction of the MUSE + AO data. 
GvdV acknowledges funding from the European Research Council (ERC) under the European Union's Horizon 2020 research and innovation programme under grant agreement No 724857 (Consolidator Grant ArcheoDyn).
E.M.C. is funded by Padua University through grants DOR1699945/16, DOR1715817/17, DOR1885254/18, and
BIRD164402/16.
RMcD is the recipient of an Australian Research Council Future Fellowship (project number FT150100333).
\end{acknowledgements}

\bibliographystyle{aa} % style aa.bst
\bibliography{References}

\begin{appendix}
\section{imfit and pPXF fit distributions}

\begin{figure}
\includegraphics[width=0.5\textwidth]{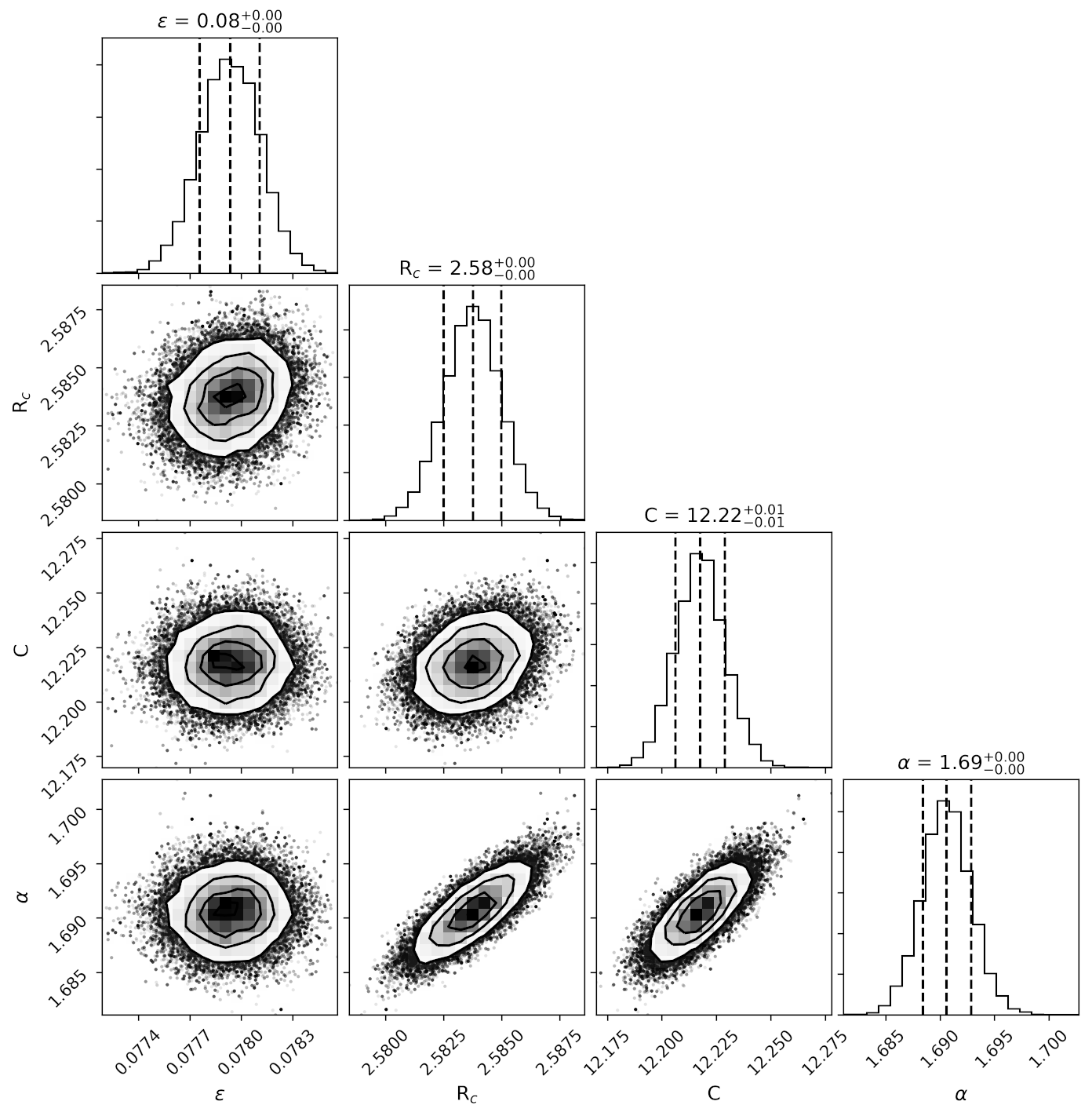}
\caption{Corner plot showing the distribution of parameters for the MCMC \textsc{imfit} fit of the single generalized King profile to the F475W data.}
\label{fig:corner}
\end{figure}

\begin{figure}
\includegraphics[width=0.5\textwidth]{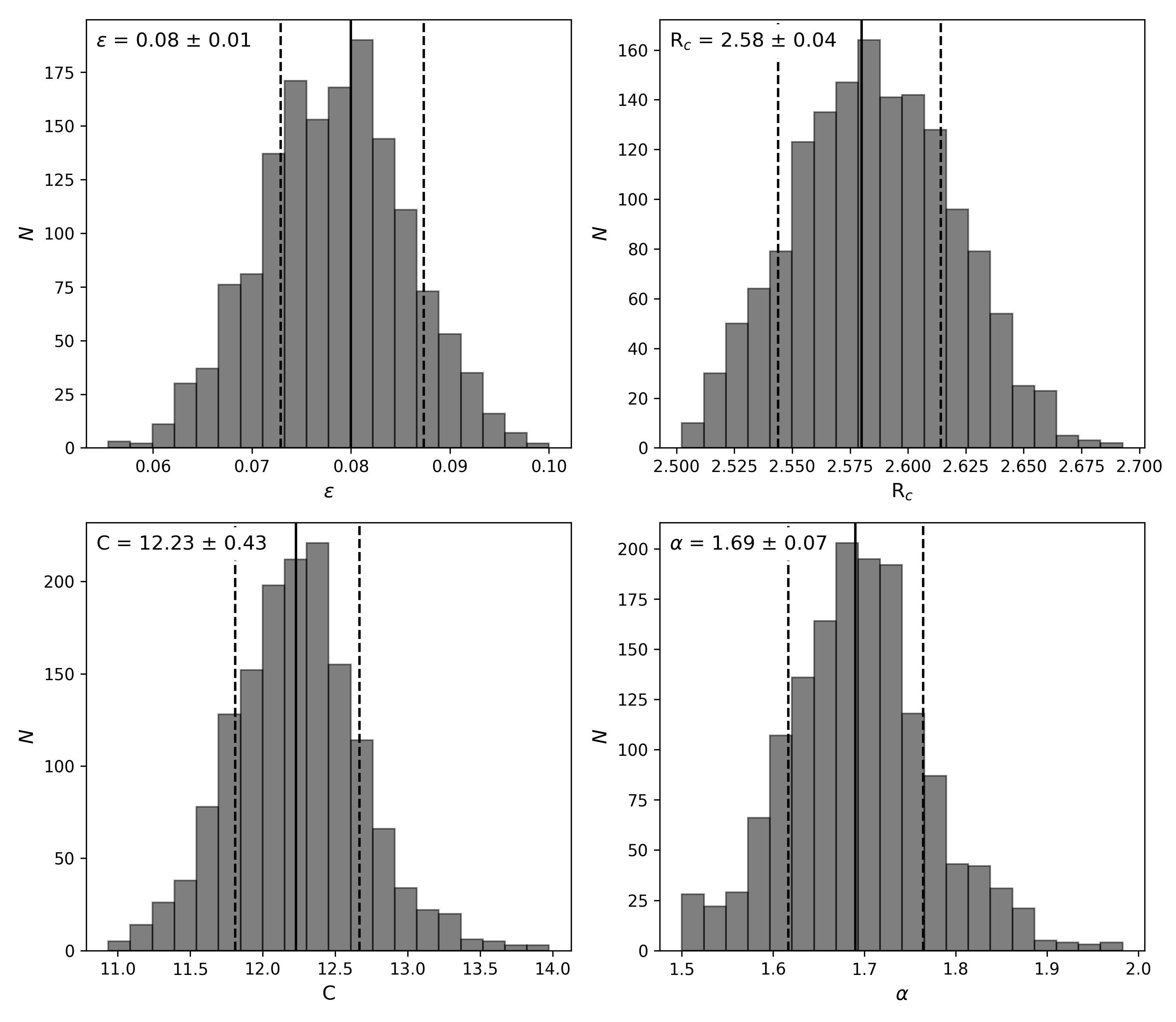}
\caption{Histograms showing the results from bootstrapping around the MCMC best-fit parameters of the single generalized King profile fit to the F475W data. We use 1000 iterations.}
\label{fig:bootstrap}
\end{figure}

We use \textsc{imfit}'s capability to perform a MCMC analysis to determine the best fitting model parameters. Figure \ref{fig:corner} shows a corner plot displaying the distribution of parameters from the MCMC \textsc{imfit} run on the F475W data with a single generalized King profile. 
Overall, the distributions are well defined and Gaussian, although there are small degeneracies between some parameters. However, the uncertainties on the best-fitting values are very small ($<$1 \%), possibly because the UCD is very bright compared to the galaxy-subtracted background and the HST data is of high quality. 
To get more realistic errors, we perform \textsc{imfit}'s bootstrapping analysis on the best-fit MCMC parameters with 1000 iterations. The resulting histograms are shown in Figure \ref{fig:bootstrap}.

\begin{figure}
    \centering
    \includegraphics[width = 0.49\textwidth]{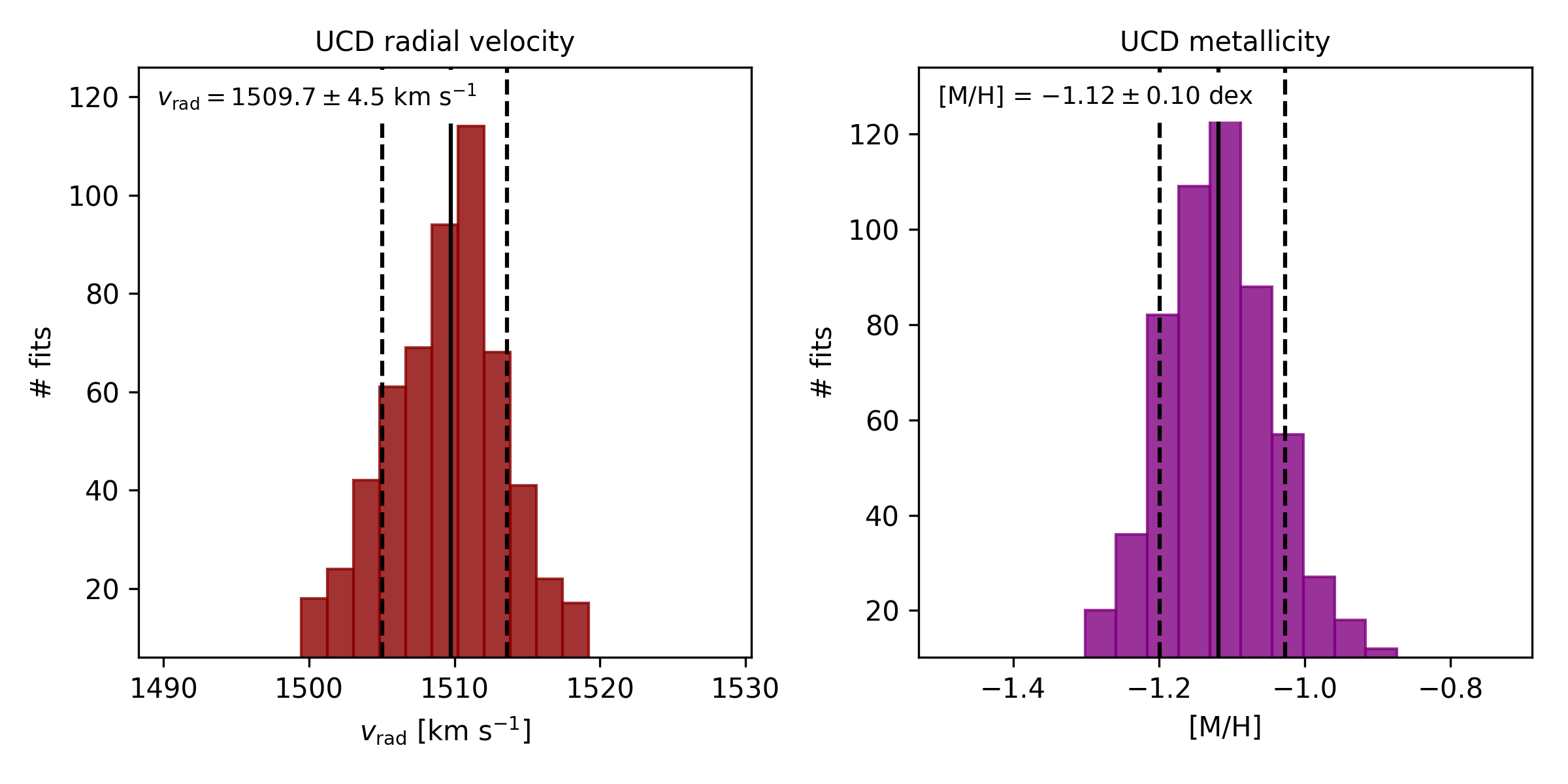}
    \caption{Radial velocity and mean metallicity distributions. These distributions were obtained by fitting 600 realisations of the UCD spectrum with \textsc{pPXF} in the full MUSE wavelength range.}
    \label{fig:vs_metals_distributions}
\end{figure}

As described in Section \ref{sect:methods}, we create 600 realizations of the UCD MUSE spectrum and fit it with \textsc{pPXF} using the E-MILES SSP library to create well-sampled distributions for the radial velocity and mean metallicity. The distributions are shown in Figure \ref{fig:vs_metals_distributions}. 

\section{Catalogue of literature UCDs}
\label{sect:UCD_catalog}
For the purpose of comparing FCC\,47-UCD1 with others, we have collected a sample of about 350 UCDs and other dense stellar systems from the literature. Our collection is taken from:
\begin{itemize}
\item \cite{Mieske2008}. The authors have published a catalogue of 55 dense stellar systems around NGC 1399, M87, CenA and others. The references of these sources are \cite{Hasegan2005, Hilker2007, Rejkuba2007, Evstigneeva2007, McLaughlin2005, Meylan2001, Barmby2007, deMarchi1999}. Most of these sources probably are UCDs, but they also include GCs around CenA.
\item \cite{Brodie2011}. Their sample includes 25 UCDs around M87.
\item \cite{Liu2015}. Their sample includes 142 UCDs around M87, M49 and M60. Some of the sources associated with M60 might truly be hosted by M59. The authors report $g$-band magnitudes and ($u - g$) colours that we convert to $V$-band using the relations from Lupton\footnote{\url{www.sdss.org/dr7/algorithms/sdssUBVRITransform.html}}.
\item \cite{NorrisKannappan2011}. The authors have discovered three UCDs around NGC 3923, one around NGC 4546 and study one around the Sombrero galaxy (M104), first discovered by \cite{Hau2009}. 
\item \cite{Price2009}. The sample contains 6 massive compact objects in the Coma galaxy cluster. Three of them are clearly identified as cE galaxies.
\item \cite{Evstigneeva2007a}. This includes three UCDs around Dorado and two around NGC 1400.
\item \cite{Misgeld2011b}. Their catalogue includes 118 members of the Hydra I galaxy cluster, 52 of which are bright enough to be considered to be UCDs. 
\item \cite{Mieske2007, Mieske2009}. The sample includes 30 compact objects found in the Centaurus galaxy cluster. Six of the objects are associated with the Cen45 subcluster dominated by NGC 4709 while the rest is associated to Cen30 (NGC 4696).
\item M31: We select the brightest ($M_V < -10.0$) GCs from the revised Bologna Catalogue \citep{Galleti2004} as well as the extended clusters found around M31 from \cite{Huxor2011}.
\end{itemize}
We further add GC-2 as a UCD around M81 \citep{Ma2017}, W3 in NGC 7252 \citep{Maraston2003}.
We remove double entries from the sample and calculate the projected distances to their assumed host galaxy if coordinates are given. For this and for the calculation of absolute magnitudes, we use distances either from the quoted works directly or from the NED database\footnote{\url{https://ned.ipac.caltech.edu/}}. We have half-light radii and absolute $V$-band magnitudes (only $B$-band magnitudes for the sources from \citealt{Price2009}) for almost all sources and have included the radial velocities for $\sim 300$ UCDs. Almost all UCDs have an assigned host from the literature. Mostly, this is the central galaxy of the field that was observed and sometimes it is the closest in projection. We do not test how likely the association is, but give for each host galaxy basic properties such as sky coordinates, magnitudes and heliocentric radial velocities. 
We show the catalogue in Table \ref{tab:UCD_cat} in a shortened form.  The full catalogue will be available online.

\begin{table*}
\centering
\caption{Catalogue of literature ultra compact dwarf galaxies in shortened form.}
\begin{threeparttable}
\begin{tabular}{c c c c c c c c c c}
\hline \hline
\multicolumn{2}{c}{Name} & RA (J2000) & DEC (J2000) & v  & $M_V$ & $r_\text{proj}$ & $r_\text{eff}$ & References \\
 Host & UCD    &             &            & [km s$^{-1}$] & [mag] & [kpc]     & [pc]      \\ \hline
  FCC\,47 (NGC\,1336) & & 03:26:32.19 & $-$35:42:48.80 & 1444.4\tnote{a} & - & - & - & NED\\
 & UCD1 & 03:26:32.92 & $-$35:42:56.09 & 1509.7 & $-$11.15 & 1.1 & 23.6 & This work\\
NGC\,1399 & & 03:38:29.03 & $-$35:27:02.36 & 1424.9 & - & - & - & NED \\
 & UCD1 & 03:37:03.30 & $-$35:38:04.60 &  - & $-$12.20 & 120.3 & 22.39 & 1, 2, 3 \\
& UCD3 & 03:38:54.10 & $-$35:33:33.60 & 1509.0 & $-$13.40 & 45.1 & 89.7 & 1 \\
& UCD4 & 03:39:35.90 & $-$35:28:24.60 & 1902.0 & $-$12.30 & 74.3 & 29.51 & 1 \\
\multicolumn{9}{c}{$\cdots$}\\

M87 (NGC\,4486) & & 12:30:49.42 & +12:23:28.04 & 1284.0 & - & - & - & NED \\
& VUCD1 & 12:30:07.57 & +12:36:31.00 & 1233.0 & $-$12.05 & 79.54 & 12.22 & 4 \\
& VUCD2 & 12:30:48.20 & +12:35:10.92 & 872.5 & $-$11.97 & 56.24 & 10.96 & 4 \\
& VUCD3 & 12:30:57.40 & +12:25:44.80 & 713.0 & $-$12.60 & 14.41 & 18.71 & 1, 5\\
& VUCD7 & 12:31:52.93 & +12:15:59.50 & 985.0 & $-$13.40 & 82.62 & 96.83 & 1, 5\\
\multicolumn{9}{c}{$\cdots$}\\

M49 (NGC\,4472) & & 12:29:46.76 & +08:00:01.71 & 980.9 & - & - & - & NED \\
& UCD1 & 12:27:08.94 & +7:42:28.31 &  - & $-$11.87\tnote{b} & 205.65 & 44.87 & 6 \\
& UCD2 & 12:27:44.26 & +7:25:42.47 & 1107.0 & $-$10.64\tnote{b} & 219.89 & 13.74 & 6 \\
& UCD3 & 12:28:03.46 & +6:53:43.11 &  - & $-$9.84\tnote{b} & 341.17 & 24.89 & 6 \\
\multicolumn{9}{c}{$\cdots$}\\

M60 (NGC\,4649)& & 12:43:39.98 & +11:33:09.74 & 1110.1  & - & - & - & NED \\
& UCD1 & 12:39:09.32 & +11:21:47.94 &  - & $-$10.25\tnote{b} & 322.92 & 12.50 & 6 \\
& UCD2 & 12:39:49.21 & +11:01:35.31 & -  & $-$10.22\tnote{b} & 310.95 & 11.64 & 6 \\
& UCD3 & 12:40:02.08 & +10:55:17.19 &  - & $-$9.87\tnote{b} & 314.328 & 13.84 & 6 \\
\multicolumn{9}{c}{$\cdots$}\\

NGC\,3923 & & 11:51:01.8 &	$-$28:48:22.0 & 1739.1 & - & - & - & NED  \\
 & UCD1 & 11:51:04.1 & $-$28:48:19.8 & 2096.9 & $-$12.43 & 3.13 & 12.3 & 7 \\
 & UCD2 & 11:50:55.9 & $-$28:48:18.4 & 1500.5 & $-$11.93 & 8.01 & 13.0 & 7 \\
 & UCD3c & 11:51:05.2 & $-$28:48:58.9 & - & $-$11.29 & 5.98 & 14.09 & 7 \\

 \multicolumn{9}{c}{\large$\cdots$}\\
\end{tabular}
\begin{tablenotes}
\item[a)] From Fahrion et. al. in prep.
\item[b)] Converted from $g$-band magnitudes and ($u$ - $g$) colours.
\item References:
1 - \cite{Mieske2008}, 2 - \cite{Hilker2007}, 3 - \cite{Drinkwater2003}, 4 - \cite{Brodie2011}, 5 -  \cite{Evstigneeva2007}, 6 - \cite{Liu2015}, 7 - \cite{NorrisKannappan2011}
\end{tablenotes}
\end{threeparttable}
\label{tab:UCD_cat}
\end{table*}

\end{appendix}
\end{document}